\documentclass[aps,longbibliography,nofootinbib,showkeys]{revtex4-1}

\usepackage{amssymb}
\usepackage{amsfonts} 
\usepackage{graphicx} 
\usepackage{amsthm} 
\usepackage{amsmath} 
\usepackage{epsfig}
\usepackage{hyperref}
\usepackage{color}
\usepackage{bbold}

\usepackage{young}
\usepackage{youngtab}
\usepackage{braket}

\usepackage{caption}
\usepackage{subcaption}
\usepackage{float}

\usepackage[utf8]{inputenc}

\def\nn{\nonumber}
\def\ic{\mathrm{i}}

\def \bc {\begin{center}}
\def \ec {\end{center}}
\def \bi {\begin{itemize}}
\def \ei {\end{itemize}}

\def \ba {\begin{array}}
\def \ea {\end{array}}

\def \bea {\begin{eqnarray}}
\def \eea {\end{eqnarray}}

\def \be {\begin{equation}}
\def \ee {\end{equation}}

\newcommand{\la}{\langle}
\newcommand{\ra}{\rangle}

\def \um {\frac{1}{2}}

\def\tr {\mathrm{tr}}

\def\bc {\bar{\beta}}

\def\cL {{\cal L}}
\def\cS {{\cal S}}

\def\nb{{\vec{n}}}
\def\zb{\mathbf{z}}

\def\E{E}
\def\rmu{\mathrm{U}}
\def\dcat{{\scriptstyle\mathrm{DCAT}}}
\def\2cat{{\scriptstyle\mathrm{2CAT}}}
\def\3cat{{\scriptstyle\mathrm{3CAT}}}

\begin{document}

\title{Information diagrams in the study of entanglement in symmetric multi-quDit systems and applications to quantum phase 
transitions  in Lipkin-Meshkov-Glick D-level atom models}

\author{Julio Guerrero}
\email{jguerrer@ujaen.es: corresponding author}
\affiliation{Department of Mathematics, University of Ja\'en, Campus Las Lagunillas s/n, 23071 Ja\'en, Spain}
\affiliation{Institute Carlos I of Theoretical and Computational Physics (iC1), University of  Granada,
Fuentenueva s/n, 18071 Granada, Spain}
\author{Alberto Mayorgas}
\email{albmayrey97@ugr.es}
\affiliation{Department of Applied Mathematics, University of  Granada,
Fuentenueva s/n, 18071 Granada, Spain}
\author{Manuel Calixto}
\email{calixto@ugr.es}
\affiliation{Department of Applied Mathematics, University of  Granada,
Fuentenueva s/n, 18071 Granada, Spain}
\affiliation{Institute Carlos I of Theoretical and Computational Physics (iC1), University of  Granada,
Fuentenueva s/n, 18071 Granada, Spain}

\date{\today}

\begin{abstract}
In this paper we pursue the use of information measures (in particular, information diagrams) for the study of entanglement in symmetric multi-quDit systems. 
We use generalizations to $\rmu(D)$ of spin $\rmu(2)$ coherent states and their adaptation to parity (multicomponent Schr\"odinger cats) and we analyse one- and two-quDit reduced density matrices.  
We use these correlation measures to characterize quantum phase transitions occurring in Lipkin-Meshkov-Glick models of $D=3$-level identical atoms and we propose the rank of the corresponding 
reduced density matrix as a discrete order parameter.
\end{abstract}


\maketitle

\section{Introduction}
\noindent Information diagrams were introduced to discuss the relation between two different information measures, like von Neumann entropy and error probability 
\cite{IEEETransactions-1994}, or von Neumann and  linear entropies \cite{ie3trans2001-HarremoesTopsoe}. In the particular case of linear ($\cL$) and von Neumann ($\cS$) entropies, 
pairs $(\cL(\rho),\cS(\rho))$ are usually plotted for any valid probability distribution $\rho$. Here $\rho$ can also represent the density matrix of a quantum system 
(or rather a vector with its eigenvalues), and this is our main interest in this paper. Special attention is paid to the boundaries of the resulting information diagram region, where the associated 
probability distributions (or density matrices) will be denoted as ``extremal''.
In Ref. \cite{PhysRevA.67.022110}, a comparison is made between both entropies in the case of two qubits (see also \cite{JMOMoyaCessa-Entropy} for the case of the ion-laser interaction).
In \cite{JPhysA.36.12255},  a detailed study of information diagrams is carried out for arbitrary pairs of entropies. There it is proved that, for certain conditions 
(satisfied by linear, von Neumann and R\'enyi entropies), the extremal density matrices are always the same. Counterexamples are given but, in general, the deviation will 
be very small and we can safely assume that these extremal density matrices have universal character.

In this paper we shall use information diagrams to obtain global qualitative information of particle  entanglement in symmetric multi-quDit systems described by 
generalized ``Schr\"odinger cat'' (multicomponent $\dcat$) states (first introduced in  \cite{Dodonovcat} as two-component, even and odd, states for an oscillator). These $\dcat$ states turn out to be 
a $\mathbb{Z}_2^{D-1}$ parity adaptation of $\rmu(D)$-spin coherent (quasi-classical) states and they have  
the structure of a quantum superposition of weakly-overlapping (macroscopically distinguishable)  coherent wave packets with interesting quantum properties.
For that purpose we make use of one- and two-quDit reduced density matrices (RDM), obtained by extracting one or two particles/atoms from a composite system of $N$ identical quDits described by 
a cat state, and tracing out the remaining system. It is well known (see \cite{PhysRevA.67.022110} and references therein) that the entropy of these RDMs provides 
information about the entanglement of the system. 
We shall plot the information diagrams associated to these RDMs and extract qualitative
information about one- and two-quDit entanglement, and also about the rank of the corresponding RDM, which also provides information on the entanglement
of the original system \cite{PRA-2000-Horodecki}.

We shall apply these  results to the characterization of  quantum phase transitions (QPT) occurring in Lipkin-Meshkov-Glick models of $3$-level identical atoms, 
complementing the results of \cite{QIP-2021-Entanglement}. In particular, we have seen that the rank of the  one- and two-quDit RDMs can be considered as a discrete order parameter 
precursor detecting the existence of QPTs.

The paper is organized as follows. Section \ref{InformationDiagrams} reviews the notion of information diagram, describing its main properties, particularly with respect to the rank. 
Section \ref{DSCS} reviews the concept of $\rmu(D)$-spin coherent states and their $\mathbb{Z}_2^{D-1}$ parity adapted version, the $\dcat$.   
In Section \ref{Entanglement} we compute one- and two-quDit RDMs for the
$\2cat$ and the $\3cat$, their Linear and von Neumann entropies, plotting them and constructing the associated information diagrams.
In Section \ref{LMGsec} we use information diagrams to provide qualitative information about QPTs in Lipkin-Meshkov-Glick (LMG) models. Section \ref{conclusions} is devoted to conclusions.

\section{Information diagrams}
\label{InformationDiagrams} 
\noindent To determine the boundaries  in information diagrams \cite{ie3trans2001-HarremoesTopsoe} we need to  show that, for two different measures of entropy (or information) 
$E_1$ and $E_2$, there are maximum and minimum possible values of $E_1$ (resp. $E_2$)  for a given value of $E_2$ (resp. $E_1$) \cite{JPhysA.36.12255}. 
That is, the region $\Delta$  given by the image of the map $\rho\mapsto (E_1(\rho),E_2(\rho))$ is a bounded set in the plane, 
where $\rho$ denotes all possible probability distributions (or density matrices) for a given dimension $d$. 

Since usual measures of entropy for density matrices are based on the trace, they are invariant under changes of basis. 
Hence, the only relevant information of a density matrix is contained  in its eigenvalues, thus in this paper we shall identify density matrices $\rho$ 
with their eigenvalues $(\lambda_1,\lambda_2,\ldots,\lambda_d)$, the order being irrelevant. Therefore, for our purposes, we can identify probability distributions 
and density matrices using a vector notation in terms of eigenvalues, referring to both of then as density matrices for short. Notwithstanding, we shall continue to treat 
density matrices as matrices in some situations.

In \cite{JPhysA.36.12255} it was proved that, under rather general assumptions on the convexity/concavity of the entropy measures,
the maximum and minimum values are always attained in two standard forms of density matrices
\begin{eqnarray}
 \rho_{\rm max}(\lambda)&=&(\lambda,\bar{\lambda},\stackrel{(d-1)}{\ldots},\bar{\lambda})\,, \qquad \bar{\lambda}=\frac{1-\lambda}{d-1}\leq \lambda \,, \qquad \lambda\in[\frac{1}{d},1), \label{rhomax1}\\
  \rho_{\rm min}^{(k)}(\lambda)&=&(\lambda,\stackrel{(k)}{\ldots},\lambda,\bar{\lambda},0,\ldots,0)\, ,\quad  \bar{\lambda}=1-k\lambda < \lambda\,,\quad \lambda\in[\frac{1}{k+1},\frac{1}{k}) \label{rhomin1}
\end{eqnarray}
respectively, where $k= 1,\ldots,d-1$.
Let us write the previous equations as (convex) sums of density matrices, that in turn can be seen as lower dimensional density matrices. For that purpose denote by $\rho_k$ the maximally mixed density matrix (or equal probabilities distribution) in dimension $k$, $\rho_k=(\frac{1}{k},\stackrel{(k)}{\ldots},\frac{1}{k})=\frac{1}{k}I_k$, where $I_k$ is the identity matrix in dimension $k$. Then we have:
\begin{eqnarray}
 \rho_{\rm max}(\epsilon)&=&(1-\epsilon)\, \rho_d  + \epsilon\, \rho_1 \oplus 0_{d-1}  \,,\qquad \epsilon\in[0,1) \label{rhomax2}\\
\rho_{\rm min}^{(k)}(\epsilon) & = & (1-\epsilon)\, \rho_k\oplus 0_{d-k} +  \epsilon\,0_k\oplus\rho_1\oplus 0_{d-1-k} \,,\quad \epsilon\in(0,\frac{1}{1+k}] \label{rhomin2}
\end{eqnarray}
where $0_k$ is the null matrix (or vector) in dimension $k$ and  $k=1,\ldots,d-1$.
The relation between $\epsilon$ and $\lambda$ is $\lambda=\frac{1}{d}-\left(1-\frac{1}{d}\right) \epsilon$ for eqns. (\ref{rhomax1},\ref{rhomax2}) and $\lambda=\frac{1-\epsilon}{k}$ for eqns. (\ref{rhomin1},\ref{rhomin2}). 

In most cases, the pair of entropies  $(\cL,\cS)$ is considered, where $\cL$  and $\cS$ denote linear  and von Neumann entropies, respectively.
We shall consider here normalized linear and von Neumann entropies, i.e:
\begin{equation}
 \cL(\rho)=\frac{d}{d-1}\left(1-\mathrm{Tr}(\rho^2)\right)\,,\qquad \cS(\rho)=-\mathrm{Tr}(\rho \log_d\rho),
\end{equation}
in such a way that both entropies range from $0$ (pure states) to $1$ (maximally mixed states). 
The values of both entropies for each family of curves are:
\begin{eqnarray}
\cL(\rho_{\rm max}(\epsilon)) &=& 1 -\epsilon ^2, \nn \\
\cS(\rho_{\rm max}(\epsilon)) &=& -(d-1) \frac{1-\epsilon}{d}  \log_d \left(\frac{1-\epsilon }{d}\right)-\left(\frac{1+(d-1) \epsilon }{d}\right) \log_d \left(\frac{1+(d-1) \epsilon}{d}\right),\quad \label{EntropiesMax}
\end{eqnarray}
and 
\begin{eqnarray}
\cL(\rho_{\rm min}^{(k)}(\epsilon)) &=& \frac{d}{d-1}\left(1-\epsilon^2 - \frac{(1-\epsilon )^2}{k}\right), \nn \\
\cS(\rho_{\rm min}^{(k)}(\epsilon)) &=& -(1-\epsilon ) \log_d (1-\epsilon )-\epsilon  \log_d (\epsilon ) +(1-\epsilon ) \log_d (k). \label{EntropiesMin}
\end{eqnarray}

\begin{figure}[h]
\begin{center}
\begin{subfigure}[h]{0.3\textwidth}
         \includegraphics[width=\textwidth]{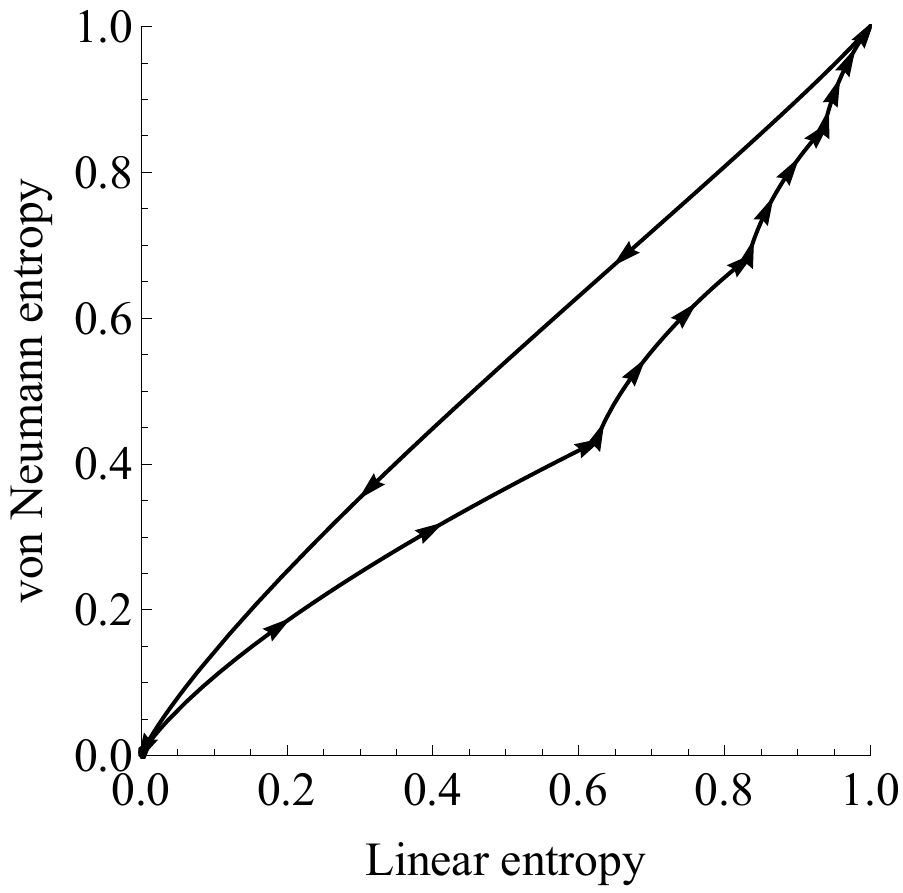}
         \caption{}
        \label{RegionDeltaa}
\end{subfigure}
\hfill
     \begin{subfigure}[h]{0.3\textwidth}
         \includegraphics[width=\textwidth]{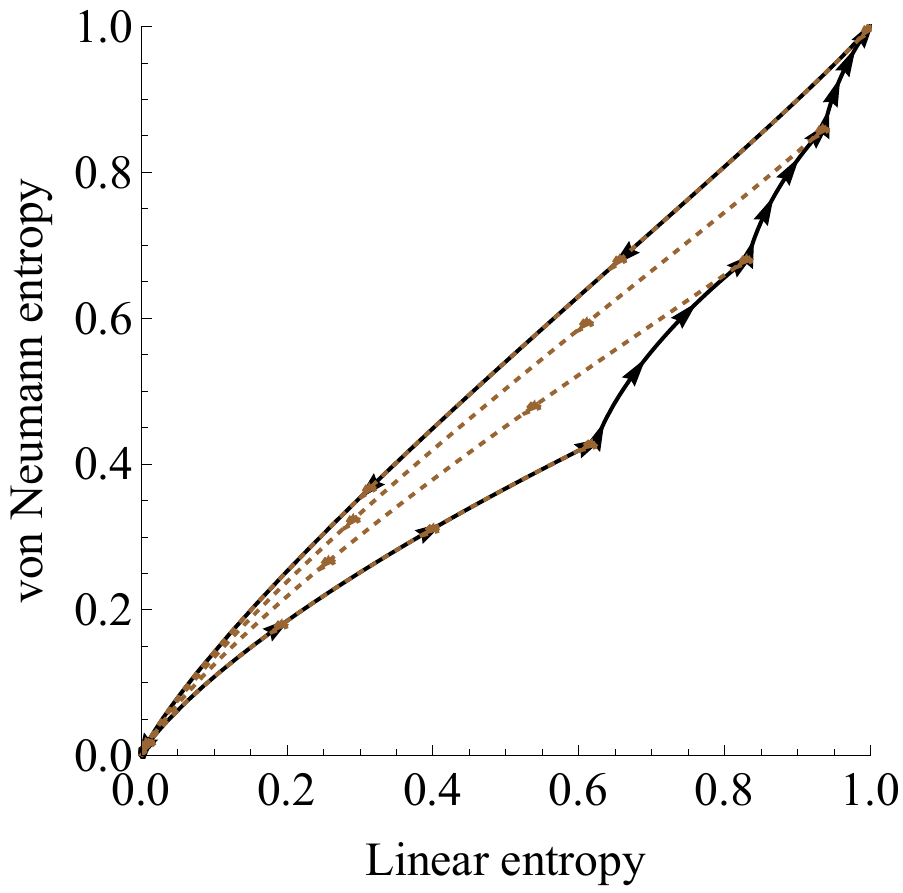}
         \caption{}
        \label{RegionDeltab}
\end{subfigure}
\hfill
    \begin{subfigure}[h]{0.3\textwidth}
         \includegraphics[width=\textwidth]{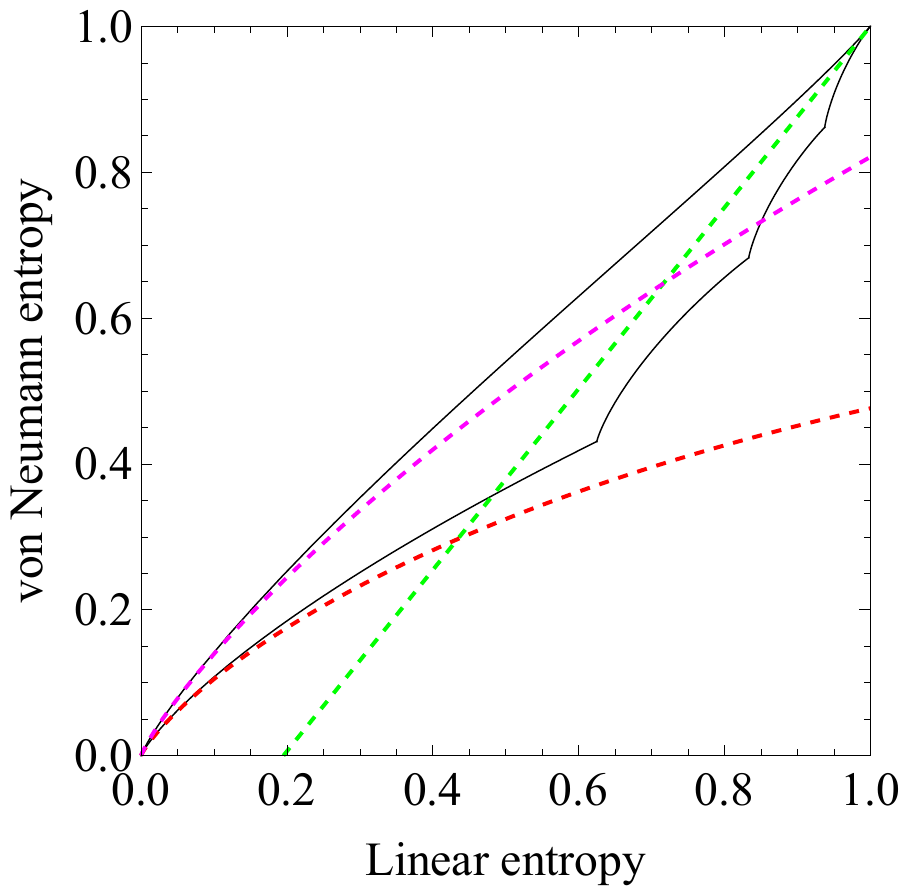}
         \caption{}
        \label{Asymptotes}
\end{subfigure}
\end{center}
\caption{(a) Information diagram for linear and von Neumann entropies in dimension $d=5$, where the region $\Delta$ is bounded by the curves associated with the density matrices 
given by eqns. (\ref{rhomax2}) (above) and (\ref{rhomin2}) (below). All curves except $\rho_{\rm max}$ are traced from left to right when $\epsilon$ increases.  (b)  Curves associated with density matrices $\bar\rho_{\rm min}^{(k)}(\epsilon)$  for $k=1,\ldots,d-1$, which are traced from right to left. Note that in the case $k=1$ the associated curve is the same as in (a), but traced backwards. Also, for $k=d-1$ the associated curve coincides with that of $\rho_{\rm max}$. (c) Plot of the asymptotic curves (\ref{asymptotic1},\ref{asymptotic2}) for density matrices near a pure state (bottom-left, red and pink, respectively)  and the asymptotic curve (\ref{asymptotic3}) near the maximally mixed state (upper-right, green)}
\label{RegionDelta}
\end{figure}

In Figure \ref{RegionDeltaa} the curves  $\rho\mapsto (\cL(\rho),\cS(\rho))$ are shown for $\rho$ equal to $\rho_{\rm max}(\epsilon)$ and $\rho_{\rm min}^{(k)}(\epsilon)$, delimiting the corresponding region $\Delta$  (we are setting $d=5$).

\noindent Note that the density matrices (\ref{rhomax2}) can be seen (for small $\epsilon$) as the maximally mixed density matrix $\rho_d$  perturbed by a rank-1 density matrix, while those of (\ref{rhomin2}) can be seen as maximally mixed density matrix of dimension $k$, $\rho_k$, perturbed by a (orthogonal) rank-1 density matrix, for $k=1,\ldots,d-1$.

It should be stressed that the range of the parameter $\epsilon$ in the curves $\rho_{\rm min}^{(k)}(\epsilon)$ can be extended to the interval $[0,1]$.
Let us denote by $\bar\rho_{\rm min}^{(k)}(\epsilon)$ the family of density matrices \eqref{rhomin2} for the range $\epsilon\in(\frac{1}{1+k},1]$. 
Their corresponding curves in the information diagram  are shown in Figure \ref{RegionDeltab}.

\subsection{Information diagrams and rank of density matrices}

\begin{figure}[h]
\begin{center}
\includegraphics[width=13cm]{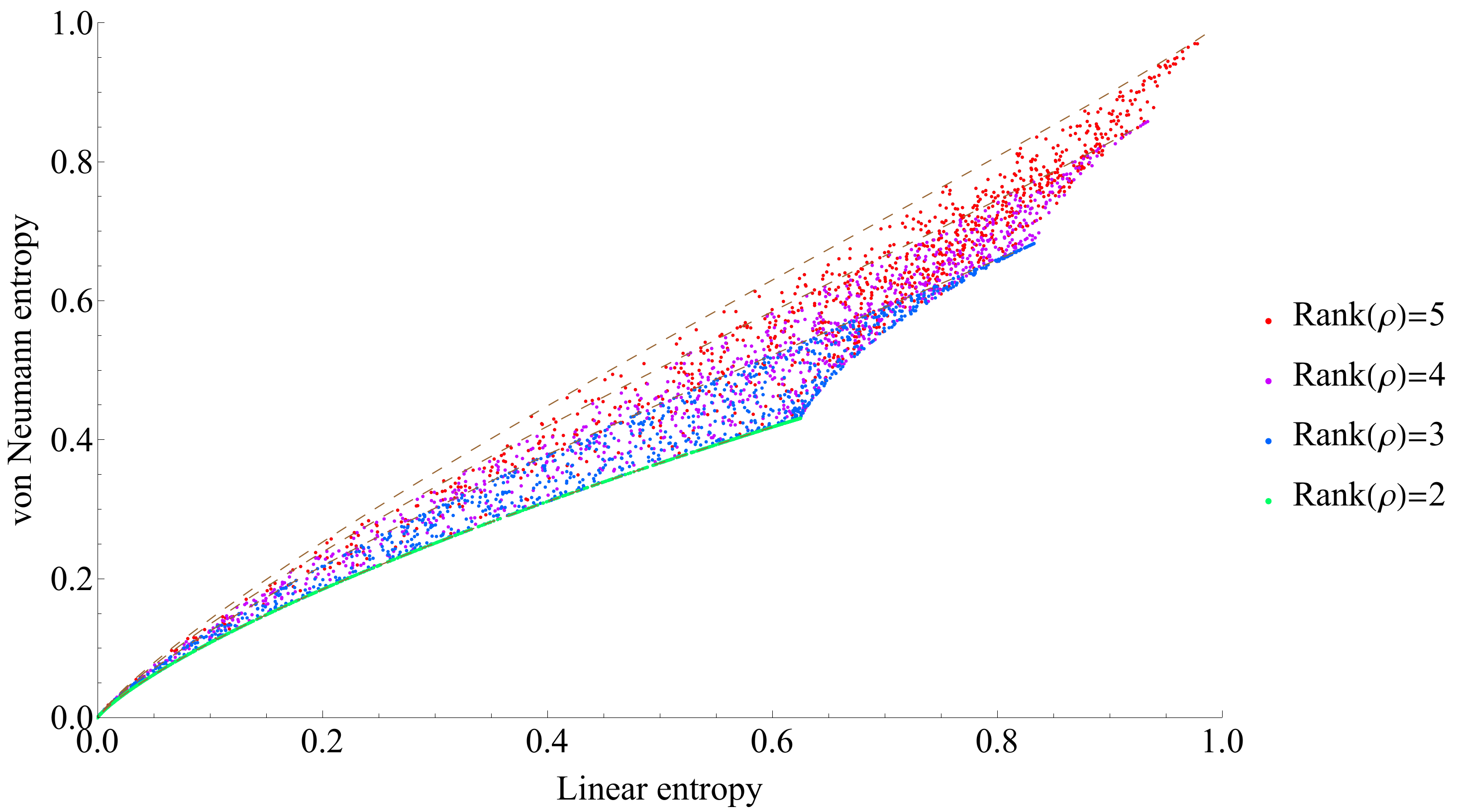}
\end{center}
\caption{Colored plot of  a sample of 20000 density matrices
of dimension $d=5$ randomly generated following a $\chi^2$ distribution for the eigenvalues in an information diagram where the different colors represent the rank of the density matrix (warmer colors represent higher ranks)}
\label{Rank}
\end{figure}

\noindent As it can be seen in Figure \ref{RegionDeltab}, there are only $d-3$ distinct $\bar\rho_{\rm min}^{(k)}$ curves, for $k=2,\ldots,d-2$. These curves divide the region $\Delta$ into $d-2$ subregions, $\Delta_k$, $k=2,\ldots,d-1$, bounded by the curves 
$\rho_{\rm min}^{(k)}$, $\bar\rho_{\rm min}^{(k)}$ and $\bar\rho_{\rm min}^{(k-1)}$.
Each subregion  $\Delta_k$  contains density matrices of  rank greater than $k$. Density matrices of rank 1 (pure states) lie on the origin, while density matrices of rank 2 lie on the curve $\rho_{\rm min}^{(1)}=\bar\rho_{\rm min}^{(1)}$. See Figure \ref{Rank} for a plot of a sample of 20000 density matrices
of dimension $d=5$ randomly generated following a $\chi^2$ distribution for the eigenvalues where the color of the corresponding point in the information diagram is associated to its rank (warmer colors correspond to higher rank).

From the expression of the extremal density matrices (\ref{rhomax1},\ref{rhomin1}), or their alternative expressions (\ref{rhomax2},\ref{rhomin2}), and the expression of the inner curves $\bar\rho_{\rm min}^{(k)}(\epsilon)$, it is clear that, for a given value of the linear entropy and a fixed rank $k+1$, the extreme values of the von Neumann entropy  are reached for $k$ identical eigenvalues. If the remaining eigenvalue is larger than the rest (i.e. we are in $\bar\rho_{\rm min}^{(k)}$) then there is a maximum, and if it is smaller than the rest (in $\rho_{\rm min}^{(k)}$) then it is a minimum of von Neumann entropy. 

\subsection{Asymptotic curves}\label{asymptoticsec}

\noindent It is interesting to obtain approximate expressions for the function $\cS(\cL)$ in some regions of the information diagram. Near a pure state (bottom left of the information diagram), we have the following asymptotic expressions for the curves $\rho_{\rm max}$ and $\rho_{\rm min}^{(1)}$:
\begin{eqnarray}
 \cS(\cL) &=&  \frac{d-1}{2 d \log (d)}\big[ \left(1 + \log (2d)\right)\cL-
 \cL\log (\cL)\big],\label{asymptotic1}\\
 \cS(\cL) &=& \frac{d-1}{2 d \log (d)} \big[  \left(1 + \log (2d)-\log(d-1)\right)\cL-\cL \log(\cL) \big], \label{asymptotic2}
\end{eqnarray}
respectively. Near the maximally mixed density matrix (upper right of the information diagram), both $\rho_{\rm max}$ and $\rho_{\rm min}^{(d-1)}$ collapse 
into the same curve, with asymptotic expression:
\begin{eqnarray}
\cS(\cL )=1-  \frac{d-1}{2 \log (d)} (1-\cL). \label{asymptotic3}
\end{eqnarray}
See Figure \ref{Asymptotes} for a plot of these asymptotic curves in an information diagram with $d=5$.

Once we have explained what the information diagrams are, and their main features, we shall use them in the study of one- and two-quDit entanglement of 
generalized Schr\"odinger cat states, which arise as a parity adaptation of  $\rmu(D)$-spin (symmetric multi-quDit) coherent states.

\section{$\rmu(D)$-spin coherent states and their adaptation to parity in symmetric multi-quDit systems}
\label{DSCS}

\noindent In this section we introduce the main ingredients and notation required to define parity adapted  $\rmu(D)$-spin coherent states in symmetric multi-quDit systems. 
These kind of states where introduced long ago in  \cite{Dodonovcat} as nonclassical (even and odd) states of light. We shall particularize to $D=2$ and $D=3$ for practical cases. 
See \cite{QIP-2021-Entanglement} for a more detailed study of the general case.

We consider a system of $N$ identical (indistinguishable) quDits, namely, $D$-level identical atoms. Denoting by $a^\dag_i$ (resp. $a_i$) the 
creation (resp. annihilation) operator of an atom  in the $i$-th level (namely, $i=1,2$ for ground and excited --or spin up and down-- in the case $D=2$, or $i=1,2,3$ 
for a 3-level atom in the case of $D=3$), the collective $\rmu(D)$-spin 
operators  can be expressed (in the fully symmetric representation) as bilinear products of creation and annihilation operators as 
(Schwinger representation)
\be
S_{ij}=a_i^\dag a_j, \quad i,j=1,\dots,D,\label{collop}
\ee
which generate the unitary symmetry $\rmu(D)$. The operator $S_{ii}$ represents the number of quDits in the level $i$, whereas $S_{ij}, i\not=j$ are raising and lowering (tunneling) operators. 
The fully symmetric representation space of $\rmu(D)$ is embedded into Fock space, with Bose-Einstein-Fock basis ($|\vec{0}\ra$ denotes the Fock vacuum)
\be
|\vec{n}\ra=|n_1,\dots, n_D\ra=
\frac{(a_1^\dag)^{n_1}\dots(a_D^\dag)^{n_D}}{\sqrt{n_1!\dots n_D!}}|\vec{0}\ra, \label{symmetricbasis}
\ee
when fixing  $n_1+\dots+n_D=N$ (the linear Casimir $C_1=S_{11}+\dots+S_{DD}$) to the total  number $N$ of quDits.  Collective $\rmu(D)$-spin operator  \eqref{collop} matrix elements are given by
\bea
\la\vec{m}|S_{ii}|\vec{n}\ra&=&n_i\delta_{\vec{m},\vec{n}}, \label{Sijmatrix}\\ 
\la\vec{m}|S_{ij}|\vec{n}\ra&=&\sqrt{(n_i+1)n_j}\delta_{m_i,n_{i}+1}\delta_{m_j,n_{j}-1}\prod_{k\not=i,j}\delta_{m_k,n_k},\; \forall i\not=j.\nonumber
\eea
The expansion of a general symmetric $N$-particle state $\psi$ in the Fock basis will be written as
\be
|\psi\ra=\sum_{\vec{n}}{}'\,c_{\vec{n}}|\vec{n}\ra=\sum_{n_1+\dots+n_D=N} c_{n_1,\dots,n_D}|n_1,\dots, n_D\ra,\label{psisym}
\ee
where $\sum'$ is a shorthand for the restricted sum. 
Among all symmetric multi-quDit states, we shall pay special attention to $\rmu(D)$-spin coherent states (DSCSs for short), which adopt the multinomial form\footnote{In eq. \eqref{cohND} and the following ones we have put $z_1=1$, where $z_1$ is the parameter multiplying $a_1^\dag$, see \cite{QIP-2021-Entanglement}. Consequently, it has been removed from the expression of $|\zb\ra$. }
\be
|\zb\ra=|z_2,\dots,z_D\ra=\frac{1}{\sqrt{N!}}\left(
\frac{a_1^\dag+z_2 a_2^\dag+\dots+z_D a_D^\dag}{\sqrt{1+|z_2|^2+\dots+|z_D|^2}}\right)^{N}|\vec{0}\ra,\label{cohND}
\ee
and are labeled by complex points $\zb=(z_2,\dots,z_D)\in \mathbb{C}^{D-1}$. These DSCSs can be seen as Bose-Einstein condensates (BECs) of $D$ modes, generalizing the spin $\rmu(2)$ (binomial) coherent states of two modes introduced by \cite{Radcliffe} and \cite{GilmorePhysRevA.6.2211} long ago.  
 If we order levels 
$i=1,\dots,D$ from lower to higher energies, the state  $|\zb=0\ra$ would be the ground state, whereas general $|\zb\ra$ could be seen as coherent excitations. Coherent states are sometimes called ``quasi-classical'' states and we shall 
see in Section  \ref{LMGsec} that $|\zb\ra$ turns out to be a good variational state that reproduces the energy and wave function of the ground state of multilevel LMG atom models in the thermodynamic (classical) limit $N\to\infty$. 

Expanding the multinomial \eqref{cohND}, we identify the coefficients $c_\nb$ of the expansion \eqref{psisym} of the DSCS $|\zb\ra$ in the Fock basis as
\be
c_\nb(\zb)=\sqrt{\frac{N!}{\prod_{i=1}^D n_i!}}\frac{\prod_{i=2}^D z_i^{n_i}}{|\zb|^{N}},\label{coefCS}
\ee
where we have written $|\zb|\equiv({\zb}\cdot\zb)^{1/2}=(1+\sum_{i=2}^D |z_i|^2)^{1/2}$ for the ``length'' of $\zb$. Note that DSCS are not orthogonal (in general) since 
\be \la \zb'|\zb\ra=\frac{({\zb}'\cdot \zb)^N}{({\zb}'\cdot \zb')^{N/2}({\zb}\cdot \zb)^{N/2}}, \quad {\zb}'\cdot \zb\equiv 1+\bar{z}_2' z_2+\dots+\bar{z}_D' z_D,\label{scprod}
\ee
is not zero, in general. However, contrary to the standard (canonical, harmonic oscillator) CSs, they can be orthogonal when ${\zb}'\cdot \zb=0$.

In \cite{QIP-2021-Entanglement} we have shown that DSCSs are separable and exhibit no quDit entanglement (although they do exhibit interlevel entanglement). In fact they can be written as a tensor 
product of 1-quDit coherent states:
\be
|\zb\ra^{(N)}=|\zb\ra_1 \otimes |\zb\ra_2\otimes  \cdots  \otimes |\zb\ra_N\,,
\ee
where we added the superscript $(N)$ to the $N$-particle coherent state \eqref{cohND}, and $|\zb\ra_i$ denotes the one-particle coherent state for the $i$-th quDit. 
Note that this state is explicitly symmetric under the interchange of quDits  and therefore there is no need to symmetrize it.

The situation changes when we deal with parity adapted DSCSs, sometimes called ``Schr\"odinger cat states'', since they are a quantum superposition of weakly-overlapping 
(macroscopically distinguishable) quasi-classical coherent wave packets. These kind of cat states arise in several  physical situations and display interesting nonclassical properties. 
The case of even parity  cat states is particularly important since they turn out to be good variational states \cite{GilmorePhysRevA.6.2211}, reproducing the energy of the ground state 
of quantum critical models in the thermodynamic limit $N\to\infty$. In \cite{QIP-2021-Entanglement}, the even  parity multi-quDit cat state $\dcat$ have been constructed for general $D$, 
and  here we shall reproduce the construction to fix  notation.

The parity operators are defined as 
\be \Pi_j=\exp(\ic\pi S_{jj}),\quad j=1,\dots,D.\label{parityop}
\ee 
Note that $\Pi_i^{-1}=\Pi_i$ and $\Pi_1\dots \Pi_D=(-1)^N$, a constraint that says that the parity group for symmetric quDits is not 
$\mathbb{Z}_2\times\stackrel{D}{\dots}\times\mathbb{Z}_2$ but $\mathbb{Z}_2\times\stackrel{D-1}{\dots}\times\mathbb{Z}_2=\mathbb{Z}_2^{D-1}$ instead. Therefore, we can discard in our discussion one of the parity operators, and we select  $\Pi_1$ (since we will use level 1 as reference level in Sec. \ref{LMGsec}).

Parity operators are conserved when the Hamiltonian scatters pairs of particles conserving the parity of the population $n_j$ in each level $j=1,\dots,D$, like in the $D$-level LMG model considered in Sec. \ref{LMGsec}. 
Using the multinomial expansion  \eqref{cohND}, it is easy to see that the effect of parity operators on symmetric DSCSs  $|\zb\ra$ is then
\be
\Pi_i|\zb\ra=\Pi_i|z_2,\dots,z_i,\dots,z_D\ra=|z_2,\dots,-z_i,\dots,z_D\ra\,,\quad i=2,\ldots,D\,. \label{parityCS}
\ee
The projector onto the even parity subspace is given by:

\be
\Pi_{\textrm{even}}=2^{1-D}\sum_{\mathbb{b}\in\{0,1\}^{D-1}} \Pi_2^{b_2}\Pi_3^{b_3}\dots \Pi_D^{b_D}\,,
\ee
where the binary string $\mathbb{b}=(b_2,\dots,b_D)\in\{0,1\}^{D-1}$ labels the elements of the parity group $\mathbb{Z}_2^{D-1}$.  We shall
also denote the symbol $\mathbb{0}$ for the string $(0,\ldots,0)$.

Let us define the even parity  generalized Schr\"odinger  cat state
\be
|\dcat\ra=\frac{1}{{\cal N}( \dcat)}\Pi_{\textrm{even}}|\zb\ra=\frac{2^{1-D}}{{\cal N}( \dcat)}\sum_{\mathbb{b}}|\zb^\mathbb{b}\ra,\label{SC}
\ee
where $|\zb^\mathbb{b}\ra\equiv|(-1)^{b_2}z_2,\dots,(-1)^{b_D}z_D\ra$ and we are using $\sum_{\mathbb{b}}$ as a shorthand for  $\sum_{\mathbb{b}\in\{0,1\}^{D-1}}$. 
The $\dcat$  is just the projection of a DSCS onto the even parity subspace. The normalization factor is given by
\be
{\cal N}( \dcat)^2=2^{1-D} \frac{ \sum_{\mathbb{b}}  L_\mathbb{b}}{L_{\mathbb{0}}}
\ee
where $L_\mathbb{b}=1+(-1)^{b_2}|z_2|^2+\cdots + (-1)^{b_D}|z_D|^2$. We shall also use the alternative notation 
$L_\sigma\equiv L_\mathbb{b}$ for $\sigma=(-1)^\mathbb{b} =((-1)^{b_2},\ldots,(-1)^{b_D})$ for convenience.

As an illustration, let us provide  the  particular expressions of $|\dcat\ra$ for $D=2$ and $D=3$. Denoting by $|\zb\ra=|z_2\ra=|\alpha\ra$ the coherent state \eqref{cohND} for $D=2$, 
the corresponding even parity $\2cat$ state is  given by
\be
|\2cat\ra=\frac{1}{2\mathcal{N}(\2cat)} \big(|\alpha\ra+|-\alpha\ra\big),
\ee
with normalization factor
\be
\mathcal{N}(\2cat)^2={\um}\left[1+\left(\frac{1-|\alpha|^2}{1+|\alpha|^2}\right)^N\right]= {\um}\frac{L_+^N + L_-^N}{L_+^N},\label{S2C}
\ee
with $L_\pm=1\pm|\alpha|^2$. Note that the overlap $\la \alpha|-\alpha\ra=(L_-/L_+)^N\stackrel{N\to\infty}{\longrightarrow} 0$ for $\alpha\neq 0$, which means that $|\alpha\ra$ 
and $|-\alpha\ra$ are macroscopically distinguishable wave packets for any $\alpha\neq 0$ (they are orthogonal for $|\alpha|=1$). 

Likewise, denoting by $|\zb\ra=|z_2,z_3\ra=|\alpha,\beta\ra$ the coherent state \eqref{cohND} for $D=3$, the  corresponding even parity 
$\3cat$s state is explicitly given by
\be
|\3cat\ra=\frac{1}{4\mathcal{N}(\3cat)}\big(|\alpha,\beta\ra+|-\alpha,\beta\ra+ |\alpha,-\beta\ra+|-\alpha,-\beta\ra\big),\label{S3C}
\ee
where 
\bea
\mathcal{N}(\3cat)^2&=&\frac{1}{4}\left[1+\frac{(1-|\alpha|^2+|\beta|^2)^N+(1+|\alpha|^2-|\beta|^2)^N+(1-|\alpha|^2-|\beta|^2)^N}{(1+|\alpha|^2+|\beta|^2)^N}\right] \nn\\
&=&\frac{1}{4}\frac{L_{++}^N+L_{-+}^N+ L_{+-}^N+ L_{--}^N}{L_{++}^N}, 
\label{S3CN}
\eea
with $L_{\sigma_1\sigma_2}= 1+\sigma_1|\alpha|^2+\sigma_2|\beta|^2$, for $\sigma_1,\sigma_2=\pm$. 
We shall use \eqref{S3C} and \eqref{S3CN} in Section \ref{LMGsec}, when discussing a LMG model of atoms with  $D=3$ levels. The  $\3cat$ state has also been used 
in $\rmu(3)$ vibron models of molecules \cite{Calixto_2012,PhysRevA.89.032126} and Dicke models of 3-level atoms interacting with a polychromatic radiation field 
\cite{PhysRevA.92.053843,L_pez_Pe_a_2015}.

\section{Entropic measures on reduced density matrices to quantify entanglement}
\label{Entanglement}

One of the most important applications of entropy measures is to quantify the entanglement of the state of a system. For that purpose
 we define several types of bipartition of the whole system, computing the corresponding RDMs and entanglement measures for symmetric multi-quDit 
states $\psi$ in terms of linear $\mathcal{L}$ and von Neumann $\mathcal{S}$ entropies. We shall focus on  one- and two-quDit
entanglement, computing the one-  and  two-particle RDMs ($\rho_1$ and $\rho_2$) for a single and a pair of particles extracted at random from a symmetric $N$-quDit state $\psi$.  
The procedure is straightforwardly extended to $\rho_M$ for an arbitrary number $M\leq N/2$ of quDits. However, as we shall see, it is not necessary to go beyond  
two particles since the two-particle RDMs provides enough information for small values of $D$. Actually, in the particular case of $D=2$, the one-particle RDM contains 
all necessary information about the entanglement of the system.

In \cite{QIP-2021-Entanglement} we gave the general expression of the one-quDit RDM of any normalized symmetric $N$-quDit state $\psi$ like  \eqref{psisym},  expressed in terms of expectation 
values of $\rmu(D)$-spin operators $S_{ij}$ as 
\be \rho_{1}^N(\psi)=\frac{1}{N}\sum_{i,j=1}^D \la\psi| S_{ji}|\psi\ra \E_{ij},\label{rho1}\ee
where  $E_{ij}$ represent $D^2$,  $D\times D$-matrices with entries $(E_{{ij}})_{lk}=\delta _{{il}}\delta _{{jk}}$ (1 in row $i$, column $j$, and 0 elsewhere).  Likewise, the two-particle RDM of 
a symmetric state $\psi$  of $N>2$ quDits is written as \cite{QIP-2021-Entanglement}
\be
\rho_2^N(\psi)=\frac{1}{N(N-1)}\sum_{i,j,k,l=1}^D \big(\la\psi| S_{ji} S_{lk}|\psi\ra-\delta_{il}\la\psi| S_{jk}|\psi\ra\big)  \E_{ij}\otimes\E_{kl}.\label{rho2}
\ee
The matrices $E_{ij}$ are the generalization to arbitrary $D$ of standard Pauli matrices for qubits ($D=2$), namely  
$\E_{12}=\sigma_+, \E_{21}=\sigma_-, \E_{11}-\E_{22}=\sigma_3$ and 
$\E_{11}+\E_{22}=\sigma_0$ (the $2\times 2$ identity  matrix).  Actually, the one- and two-qubit RDMs for $D=2$ were already considered time ago by Wang and M\o{}lmer in \cite{Molmer}.
Here we shall consider both cases, $D=2$ (qubits) and $D=3$ (qutrits), in order to discuss the similitudes and differences.

\subsection{One-quDit reduced density matrices}

\noindent For the case of a DSCS $|\zb\ra$, the linear and von Neumann entropies of $\rho_{1}(\zb)$  are zero, i.e. there is no entanglement between quDits in a DSCS. This is because a 
DSCS is eventually obtained by rotating each quDit individually. The situation 
changes when we deal with parity adapted DSCSs or ``Schr\"odinger cat states'' like \eqref{S2C}-\eqref{S3C}. Indeed, the one-quDit RDM $\rho_1(\dcat)$ does not correspond 
now to a pure state since it has the expression (we provide its eigenvalues) 
\be
\rho_1^N(\2cat) = \frac{1}{2L_+^{N}\mathcal{N}(\2cat)^2}\left( L_+^{N-1}+L_-^{N-1}\,,\,|\alpha|^2\left(L_+^{N-1}-L_-^{N-1}\right) \right),
\ee
for an $N$-qubit system and 
\bea
\rho_1^N(\3cat) &=&\frac{1}{4L_{++}^{N}\mathcal{N}(\3cat)^2}\left( L_{++}^{N-1}+L_{-+}^{N-1}+L_{+-}^{N-1}+L_{--}^{N-1} \right. ,  \label{1RDM} \\
& &  |\alpha|^2\left(L_{++}^{N-1}-L_{-+}^{N-1}+L_{+-}^{N-1}-L_{--}^{N-1}\right) \,,\nn\\  
& & \left.|\beta|^2 \left(L_{++}^{N-1}+L_{-+}^{N-1}-L_{+-}^{N-1}-L_{--}^{N-1}\right) \right), \nn
\eea
for an $N$-qutrit system. 
Note that, for $\alpha\neq 0$ in the case of $\rho_1^N(\2cat)$, and $\alpha\neq 0$ or $\beta\neq 0$ in the case $\rho_1^N(\3cat)$, the corresponding one-quDit RDM  has rank greater that 1.
That is, unlike $|\zb\ra$, the Schr\"odinger cat $|\dcat\ra$ is not separable in the tensor product Hilbert space  $[\mathbb{C}^D]^{\otimes N}$. 
In addition, $\rho_1^N(\3cat)$ has rank 2 if $\alpha\neq 0$ or $\beta\neq 0$ and has rank 3 if both are different from zero. See below for a more detailed discussion on this point.

Since the main features of these density matrices are captured in the $N\rightarrow\infty$ (thermodynamic) limit (infinite number of quDits), 
we shall restrict ourselves to this limit, where the expression
of the (diagonalized) density matrices are simpler:
\bea
 \rho_1^\infty(\2cat) &=&\frac{1}{1+|\alpha|^2} \left(1,|\alpha|^2\right), \label{1RDM-Ninf1} \\
 \rho_1^\infty(\3cat) &=& \frac{1}{1+|\alpha|^2+|\beta|^2} \left(1,|\alpha|^2,|\beta|^2\right) \,.\label{1RDM-Ninf2}
\eea
It will be interesting to discuss also the case $|\alpha|= 1$, for qubits, and $(|\alpha|,|\beta|)=(1,1)$, for qutrits, since these values will appear as limiting points of the 
stationary curve  $(\alpha_0(\lambda),\beta_0(\lambda))$ in Eq.   \eqref{critalphabeta} for high $\lambda$, i.e. $(\alpha_0(\lambda),\beta_0(\lambda))\stackrel{\lambda\to\infty}{\longrightarrow} (1,1)$, where $\lambda$ is the strength of two-body (two-quDit) interactions in a $D$-level atom LMG model (see later in Section \ref{LMGsec}  for more information). Therefore, we are also interested in the ``high coupling limit'' 
\bea
\lim_{|\alpha|\rightarrow 1} \rho_1^\infty(\2cat) &=& \left(\frac{1}{2},\frac{1}{2}\right), \label{1RDM-alfabeta1} \\
\lim_{(|\alpha|,|\beta|)\rightarrow (1,1)} \rho_1^\infty(\3cat) &=&  
 \left(\frac{1}{3},\frac{1}{3},\frac{1}{3}\right).
\eea
Hence,  in this high coupling limit, the 1-quDit  RDM is maximally mixed and therefore the entanglement is maximum.

For $D=2$, the asymptotic behavior of $\rho_1^\infty$ for large  $|\alpha|$ is:
\be
\rho_1^\infty(\2cat) = (0,1)+O(\frac{1}{|\alpha|^2})(1,1), \;\; |\alpha|^2\gg 1,\label{1RDM-alfainf}
\ee
while for $D=3$ the limit $(|\alpha|,|\beta|)\rightarrow (\infty,\infty)$ does not exist. Actually, the asymptotic behavior of $\rho_1^\infty(\3cat)$ along  the  lines $|\alpha|=r \cos\theta, |\beta|= r\sin\theta$, , for large $r$, is:
\be
\rho_1^\infty(\3cat) = \left(0,\sin^2\theta,\cos^2\theta\right)+ O(\frac{1}{r^2})(1,1,1) \,,\;\; r\gg 1,
\label{1RDM-alfabetainf}
\ee
implying that, in this limit, the 1-quDit RDMs have in general lower ranks, exhibiting no entanglement  for $D=2$ and $D=3$ for vertical ($\theta=\pi/2$) and horizontal ($\theta=0$) directional limits.

\begin{figure}[h]
\begin{center}
\begin{subfigure}[h]{0.275\textwidth}
 \centering
 \includegraphics[width=\textwidth]{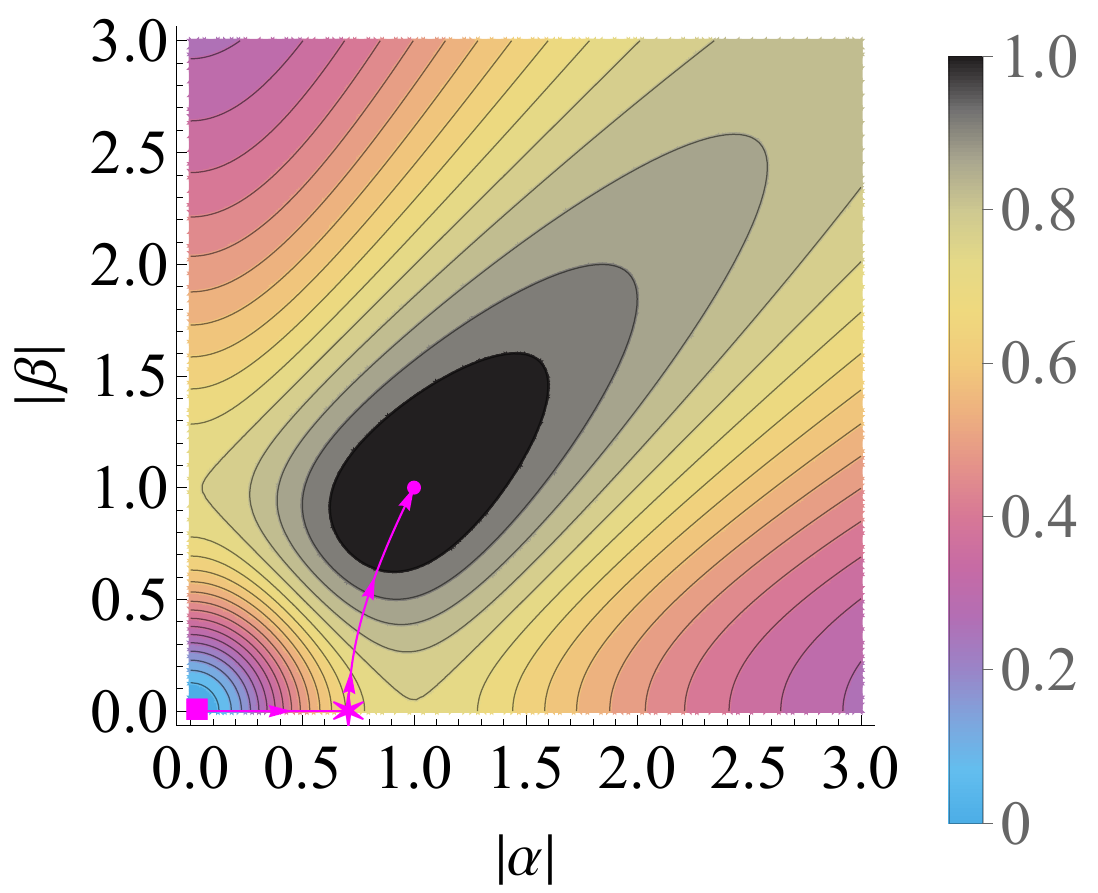}
 \caption{}
\label{purityOne3CATa}
\end{subfigure}
\begin{subfigure}[h]{0.275\textwidth}
 \centering
 \includegraphics[width=\textwidth]{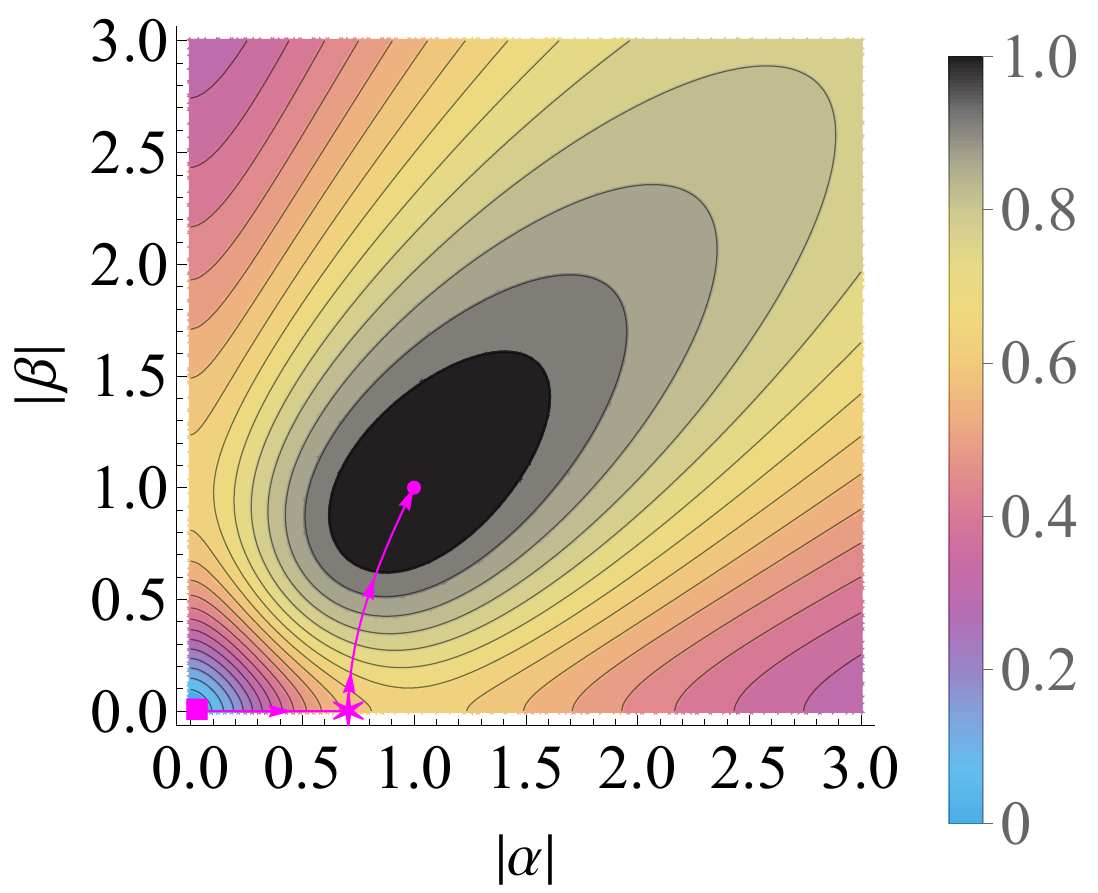}
 \caption{}
 \label{purityOne3CATb}
\end{subfigure}
\\
\begin{subfigure}[h]{0.275\textwidth}
 \centering
 \includegraphics[width=\textwidth]{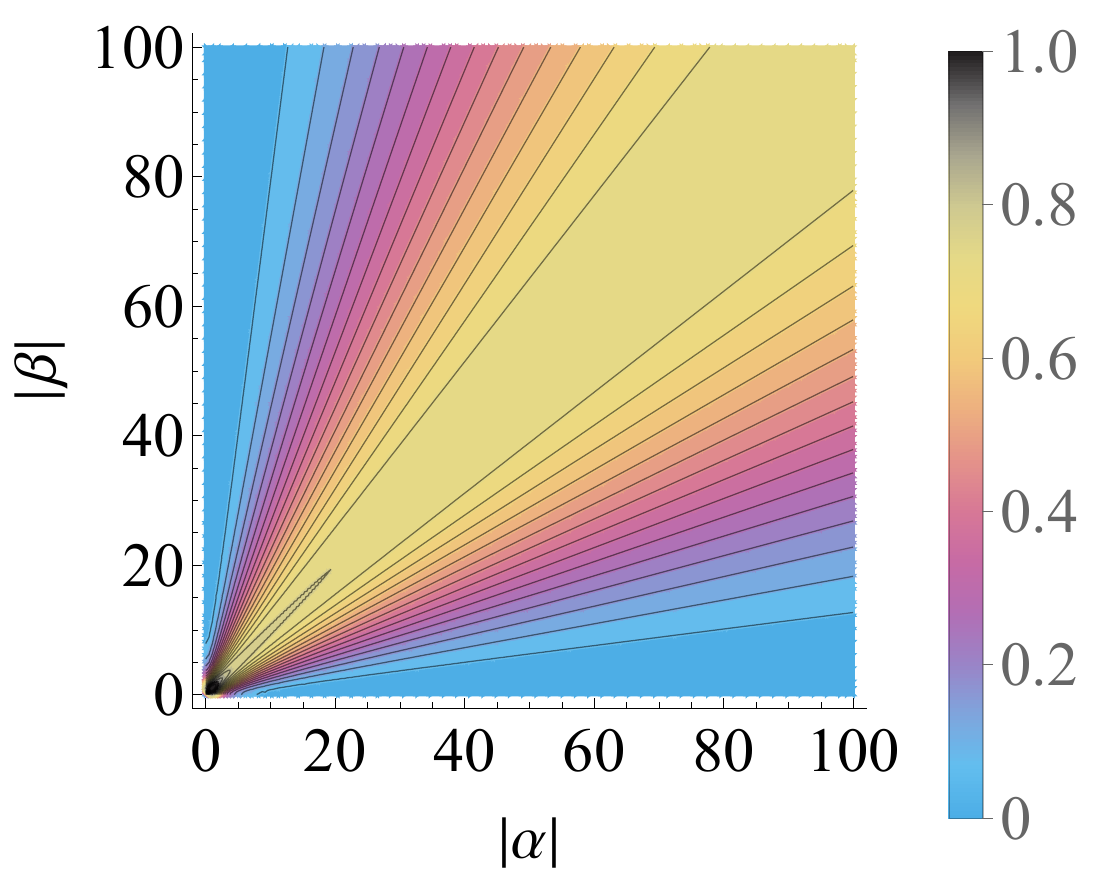}
 \caption{}
\label{purityOne3CATc}
\end{subfigure}
\begin{subfigure}[h]{0.275\textwidth}
 \centering
 \includegraphics[width=\textwidth]{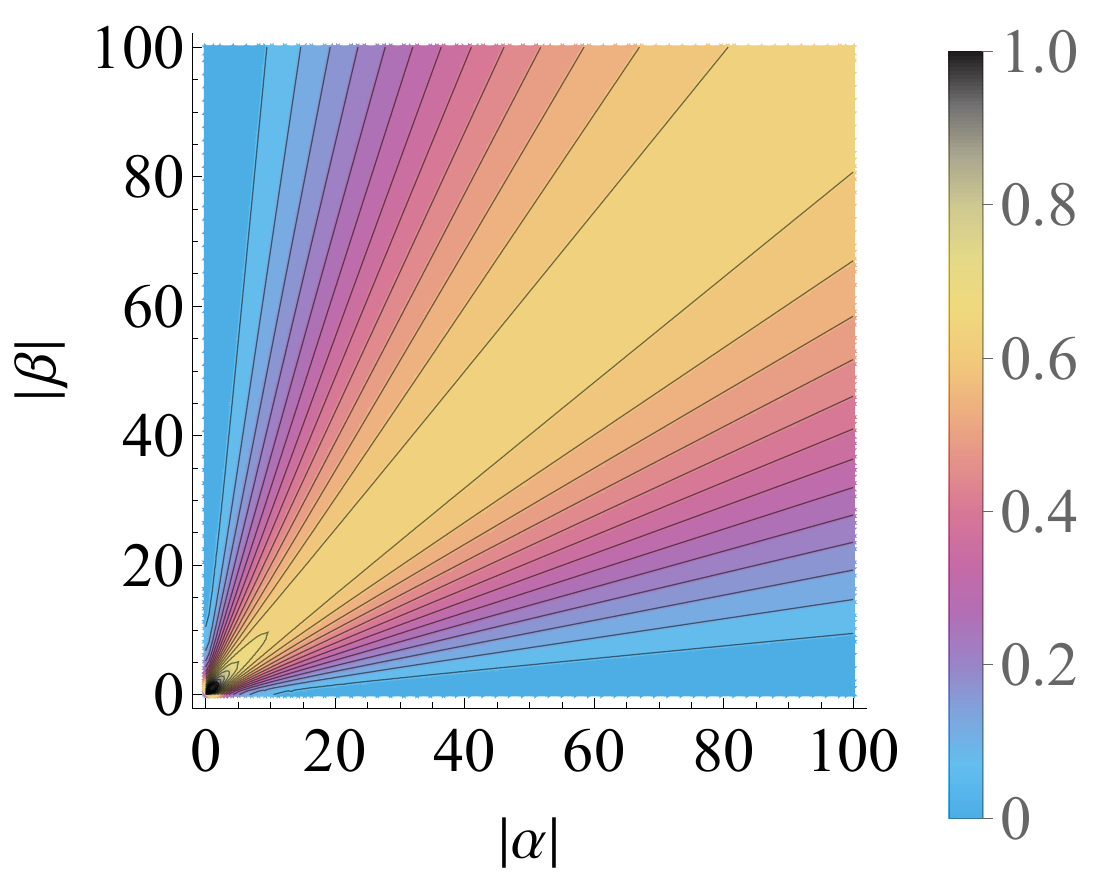}
 \caption{}
 \label{purityOne3CATd}
\end{subfigure}
\end{center}
\caption{Contour plots of (a) linear $\mathcal{L}_1^\infty$  and (b) von Neumann $\mathcal{S}_1^\infty$  entanglement entropies of the one-qutrit RDM $\rho_1^\infty(\3cat)$ of 
a $\rmu(3)$ Schr\"odinger cat \eqref{S3C} in the limit of an infinite number of qutrits, as a function  of the phase-space coordinates $\alpha, \beta$ (they just depend on moduli).
 The asymptotic behaviour of (c) $\mathcal{L}_1^\infty$ and (d)  $\mathcal{S}_1^\infty$  for large values of $|\alpha|$ and $|\beta|$ displays isentropic curves $\theta=$constant, according to the expression of 
$\rho_2^\infty(\3cat)$ in Eq. \eqref{1RDM-alfabetainf}.}
\label{purityOne3CAT}
\end{figure}

In Figures \ref{purityOne3CATa}  and \ref{purityOne3CATb} we 
represent contour  plots of linear  and von Neumann 
\be
\mathcal{L}_1^\infty=\frac{D}{D-1}(1-\tr((\rho_1^\infty)^2)),\quad \mathcal{S}_1^\infty=-\tr(\rho_{1}^\infty\log_D \rho_{1}^\infty)
\ee
entanglement entropies in the limit $N\rightarrow\infty$ of the one-qutrit RDM, $\rho_1^\infty(\3cat)$, of the  $\3cat$ in Eq. \eqref{S3C}, as a function 
of the phase-space $\mathbb CP^{2}$ coordinates $(\alpha, \beta)$ [actually, they just depend on the moduli]. Both entropies are again normalized to 1.  They attain their maximum 
value of 1 at the phase-space point $(\alpha,\beta)=(1,1)$ corresponding to a maximally mixed RDM. This behavior of the entropies, and therefore of entanglement (together with squeezing, see \cite{QIP-2021-Entanglement}) parallels that of the standard harmonic oscillator cat states where the maximun entanglement and squeezing takes place for relatively small values of the coherent state parameter \cite{Buzek92}. The diference here in the $D=3$ case is that for large values of the parameter there can still be entanglement (and squeezing), dependent on the angle of the directional limit (see eq. (\ref{1RDM-alfabetainf})). These Figures  also show (in magenta color) the values of the entropies along 
the stationary curve  $(\alpha(\lambda),\beta(\lambda))$ in Eq. \eqref{critalphabeta}, that we already mentioned before the Eq. \eqref{1RDM-alfabeta1}. 
For high interactions $\lambda\to\infty$  we have $(\alpha(\lambda),\beta(\lambda))\to (1,1)$, which means that highly coupled quDits are maximally entangled in a cat-like ground state 
(we shall come back again to this discussion later in Section \ref{LMGsec}). In Figures \ref{purityOne3CATc}  and \ref{purityOne3CATd} the asymptotic behavior for large $|\alpha|$ and $\beta|$ 
is shown, where contours of linear and von Neumann entropies coincide with the (isentropic) lines $\theta=$constant, according to the asymptotic behavior of 
$\rho_1^\infty(\3cat)$ in \eqref{1RDM-alfabetainf}.

\begin{figure}[h]
\begin{center}
\begin{subfigure}[h]{0.32\textwidth}
\centering
\includegraphics[width=\textwidth]{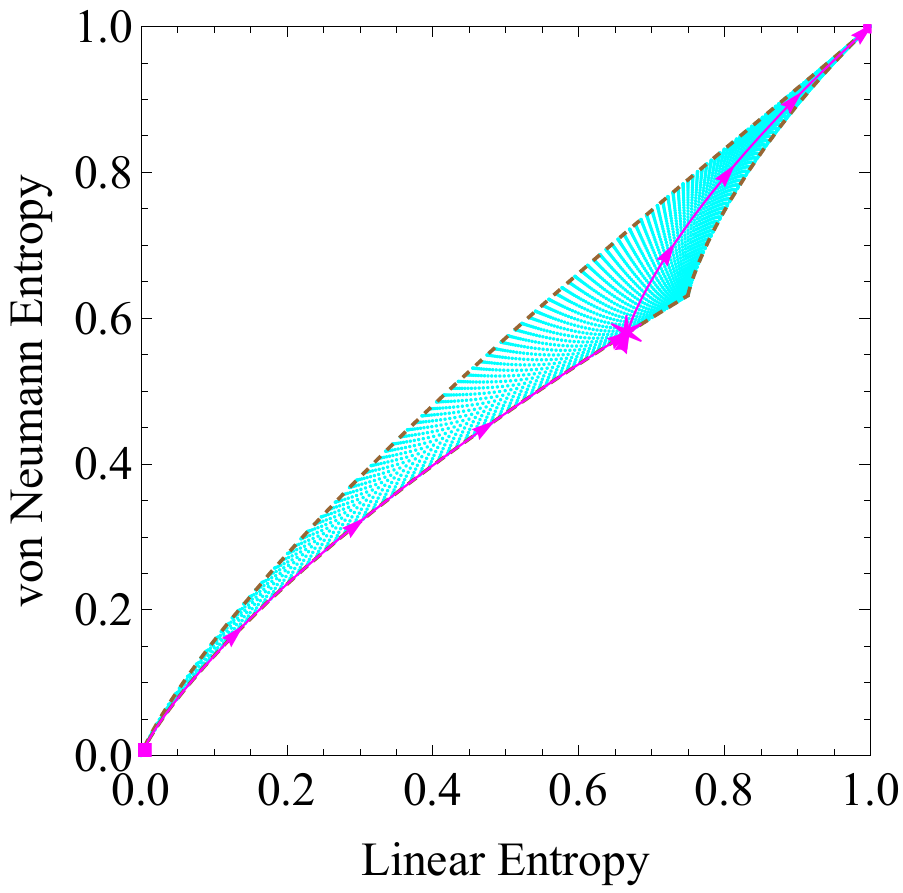}
\caption{}
\label{InfDiag-1RDM}
\end{subfigure}
\hspace{2cm}
\begin{subfigure}[h]{0.32\textwidth}
\centering
\includegraphics[width=\textwidth]{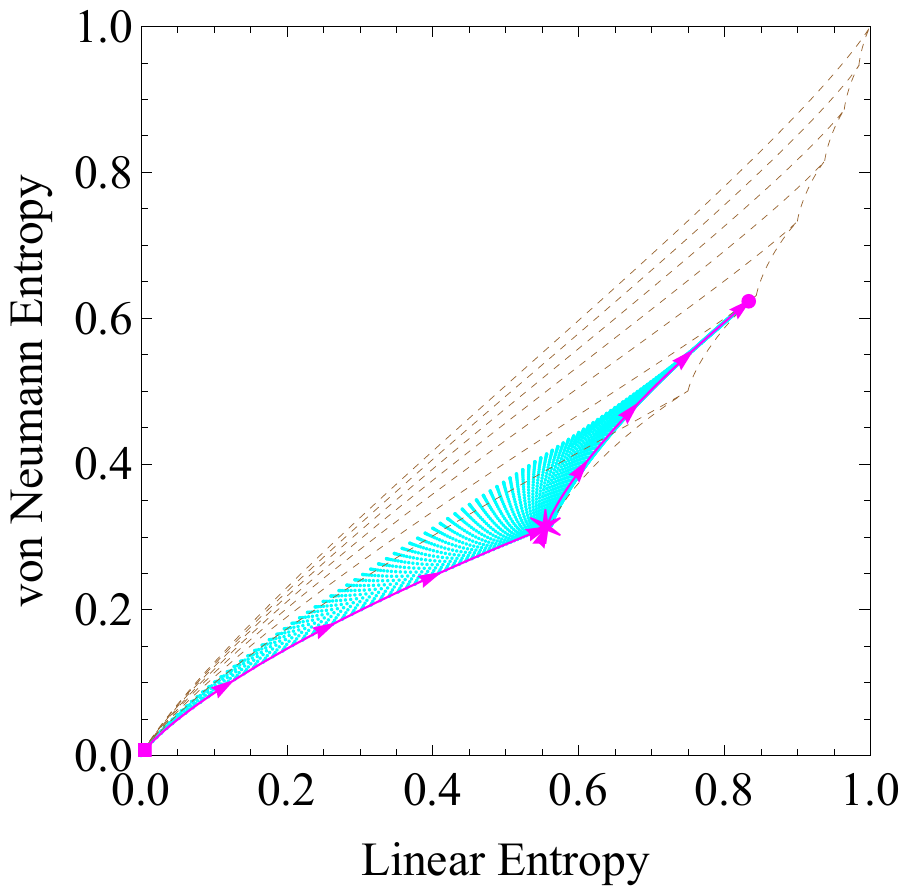}
\caption{}
\label{InfDiag-2RDM}
\end{subfigure}
\end{center}
\caption{Information diagram for the family of (a) 1-qutrit RDMs  and (b) 2-qutrit RDMs  for $\3cat$  in the limit $N\rightarrow\infty$, for all values of $|\alpha|$ and $|\beta|$. See the main text for explanation.}
\label{InfDiag-RDM}
\end{figure}

In Figure \ref{InfDiag-1RDM} we plot the information diagram for the family of 1-qutrit RDMs  for a $\3cat$  (\ref{1RDM}) in the limit $N\rightarrow\infty$, for all values of $|\alpha|$ and $|\beta|$.  It can be seen that they 
fill completely the region $\Delta$. Also, the stationary curve \eqref{critalphabeta} is shown, starting at the origin (zero entropy and therefore no entanglement), moving on the curve $\rho_{\rm min}^{(1)}$ and through the region $\Delta_2$, to finish at the maximally mixed state, indicating that this state is maximally entangled.

\subsection{Two-quDit reduced density matrices}

As for the one-quDit RDM,  the linear and von Neumann entropies for a two-quDit RDM of a DSCS $|\zb\ra$ are zero, i.e. there is no pairwise quDit entanglement in a DSCS. 
The situation changes for parity adapted DSCSs or ``Schr\"odinger cat states'' $|\dcat\ra$ like the ones in  \eqref{S2C}-\eqref{S3C}, where the two-quDit RDM $\rho_2(\dcat)$ 
in \eqref{rho2} has the expression (once diagonalized) for $D=2$ and $D=3$ and $N$ particles:

\bea
\rho_2^N(\2cat) &=& \frac{1}{2 L_+^N\mathcal{N}(\2cat)^2}  \left((1+|\alpha|^4) \left(L_+^{N-2}+L_-^{N-2}\right),2 |\alpha|^2 \left(L_+^{N-2}-L_-^{N-2}\right),0,0\right), \nn\\
\rho_2^N(\3cat) &=& \frac{1}{4L_{++}^{N}\mathcal{N}(\3cat)^2}\left((1+|\alpha|^4+|\beta|^4) \left(L_{++}^{N-2}+L_{-+}^{N-2}+L_{+-}^{N-2}+L_{--}^{N-2}\right),\right.\nn\\
& & 2 |\alpha|^2  \left(L_{++}^{N-2}-L_{-+}^{N-2}+L_{+-}^{N-2}-L_{--}^{N-2}\right), \\
& & 2 |\beta|^2  \left(L_{++}^{N-2}+L_{-+}^{N-2}-L_{+-}^{N-2}-L_{--}^{N-2}\right),\nn\\
& & 2\left. |\alpha|^2 |\beta|^2  \left(L_{++}^{N-2}-L_{-+}^{N-2}-L_{+-}^{N-2}+L_{--}^{N-2}\right),0,0,0,0,0\right) \nn
\eea
As it is deduced from the previous expressions, the 2-qudit RDM has rank 1 for $\alpha=0$ in the case of the $\2cat$, or $\alpha=\beta=0$ for the case of the $\3cat$. For $\alpha\neq 0$ (or $\beta\neq 0$ for the $\3cat$) the rank is two, and for $\alpha\neq 0$ and $\beta\neq 0$ the  rank of $\rho_2(\3cat)$ is 4. 

As for the one-quDit RDM case, it is convenient to consider the thermodynamic limit  $N\rightarrow\infty$ to obtain simpler expressions without losing important qualitative information:
\bea
 \rho_2^\infty(\2cat) &=& \frac{1}{\left(1+|\alpha|^2\right)^2}\left(1+|\alpha|^4,2 |\alpha|^2,0,0\right),  \\
 \rho_2^\infty(\3cat) &=& \frac{1}{\left(1+|\alpha|^2+|\beta|^2\right)^2} \left(1+|\alpha|^4+|\beta|^4,2 |\alpha|^2,2 |\beta|^2,2 |\alpha|^2 |\beta|^2,0,0,0,0,0 \right)  \,. \nn\label{2RDM-Ninf}
\eea
The high coupling limit, $|\alpha| \rightarrow 1$ or $(|\alpha|,|\beta|)\rightarrow (1,1)$, discussed before \eqref{1RDM-alfabeta1} for 1-quDit RDMs, looks like this now for 2-quDit RDMs:
\bea
\lim_{|\alpha|\rightarrow 1} \rho_2^\infty(\2cat) &=& \left(\frac{1}{2},\frac{1}{2},0,0\right), \label{2RDM-alfabeta1} \\
\lim_{(|\alpha|,|\beta|)\rightarrow (1,1)} \rho_2^\infty(\3cat) &=&  
 \left(\frac{1}{3},\frac{2}{9},\frac{2}{9},\frac{2}{9},0,0,0,0,0\right), 
 \eea
thus implying that, in the high coupling limit,  the 2-qubit ($D=2$) RDM   is maximally mixed of rank 2, but it doesn't attain the maximum value of the entropies.  
For $D=3$, the  2-qutrit  RDM is not even  maximally mixed of rank 4 (in fact it lies on the curve   
$\bar\rho_{\rm min}^{(3)}$), 
although the value of the entropies is very similar to that of $\rho_4$ (see Figure \ref{RegionDelta}).

For $D=2$ the asymptotic behaviour of $\rho_2^\infty$ for large  $|\alpha|$ is:
\be
\rho_2^\infty(\2cat) =(0,1,0,0)+ O(\frac{1}{|\alpha|^2})(1,1,0,0),\;\; |\alpha|\gg 1, \label{2RDM-alfainf}
\ee
while for $D=3$ the limit $(|\alpha|,|\beta|)\rightarrow (\infty,\infty)$ does not exist. The asymptotic behavior, for large $r$, along  the  lines $|\alpha|=r \cos\theta, |\beta|= r\sin\theta$, is:
\bea
\rho_2^\infty(\3cat) &=& 
\left(\frac{1}{4} (\cos (4 \theta )+3),0,0,2 \cos ^2(\theta ) \sin ^2(\theta ),0,0,0,0,0\right)\nn\\ 
& &+O(\frac{1}{r^2})(1,1,1,1,0,0,0,0,0),\;\; r\gg 1,
 \,,
\label{2RDM-alfabetainf}
\eea
implying that, in this limit, the 2-quDit RDMs have in general lower ranks, exhibiting no pairwise entanglement for $D=2$ and $D=3$ for vertical $(\theta=\pi/2)$ and horizontal $(\theta=0)$ directional limits.

\begin{figure}[h]
\begin{center}
\begin{subfigure}[h]{0.275\textwidth}
 \centering
 \includegraphics[width=\textwidth]{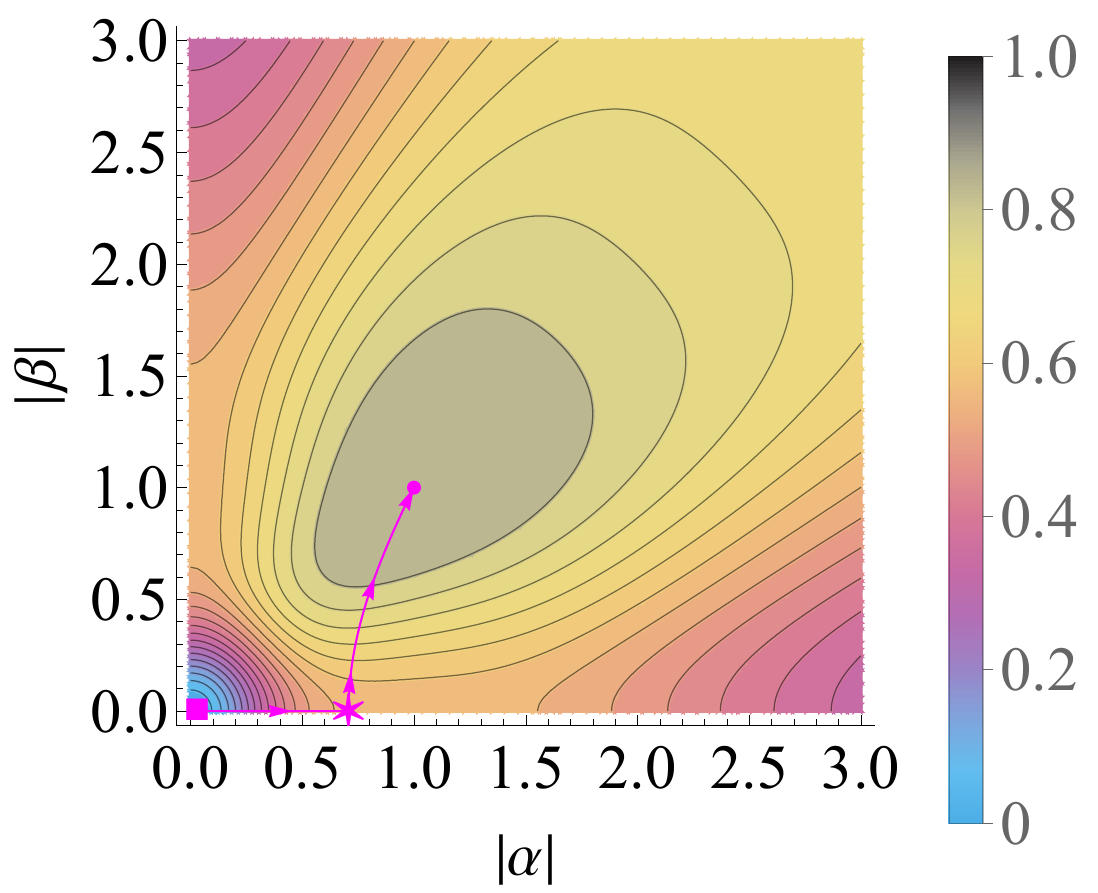}
 \caption{}
\label{purityTwo3CATa}
\end{subfigure}
\begin{subfigure}[h]{0.275\textwidth}
 \centering
 \includegraphics[width=\textwidth]{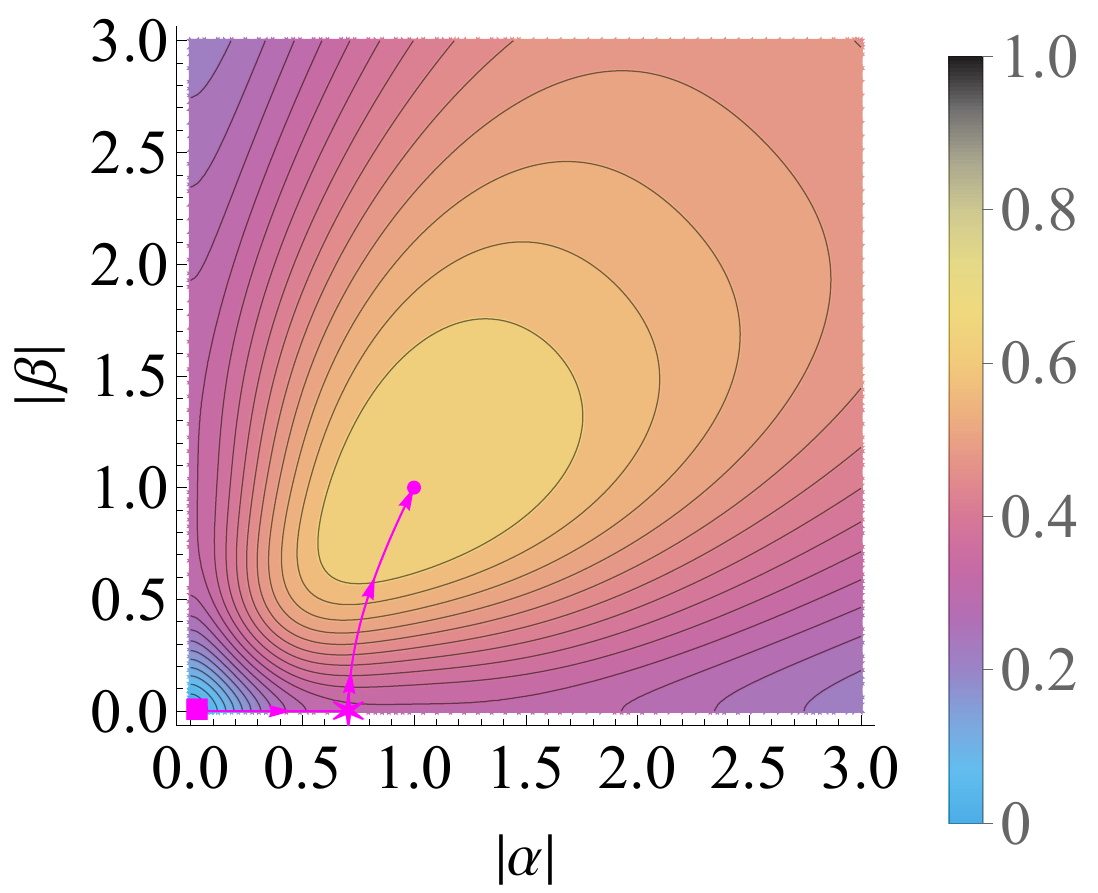}
 \caption{}
\label{purityTwo3CATb}
\end{subfigure}
\\
\begin{subfigure}[h]{0.275\textwidth}
 \centering
 \includegraphics[width=\textwidth]{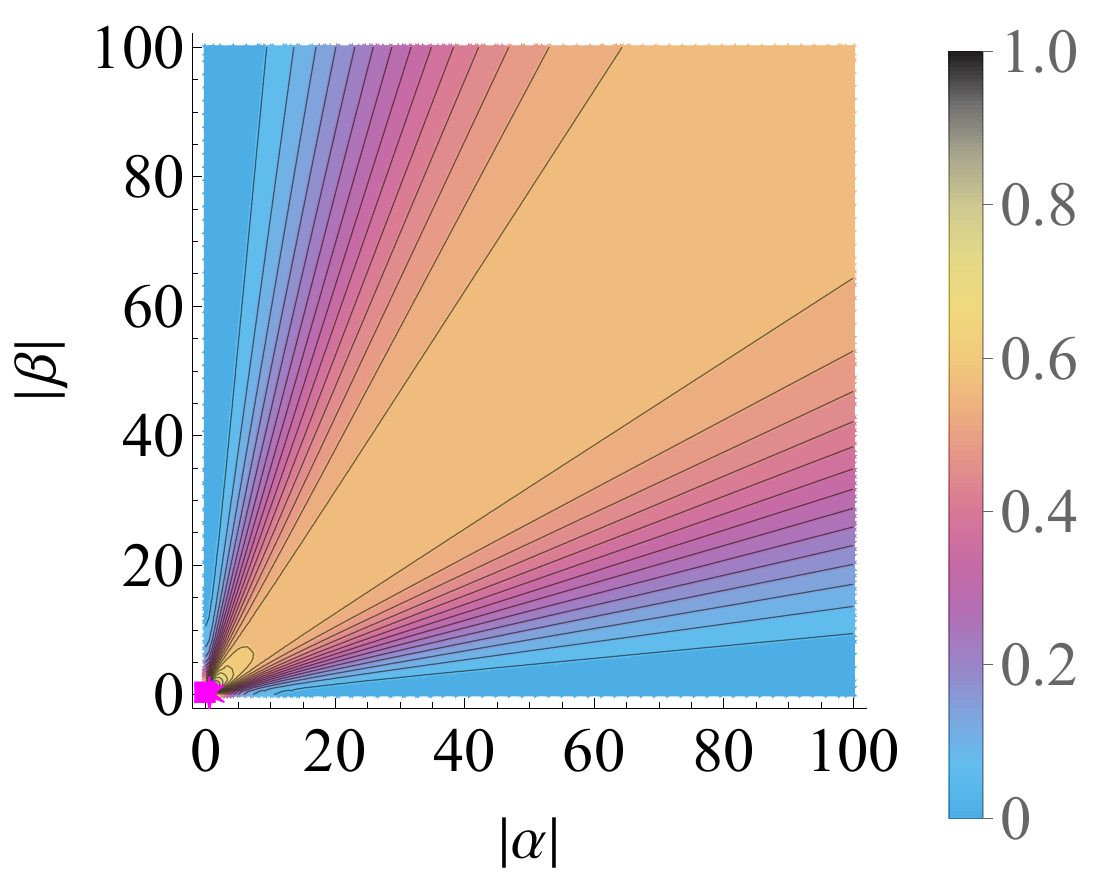}
 \caption{}
\label{purityTwo3CATc}
\end{subfigure}
\begin{subfigure}[h]{0.275\textwidth}
 \centering
 \includegraphics[width=\textwidth]{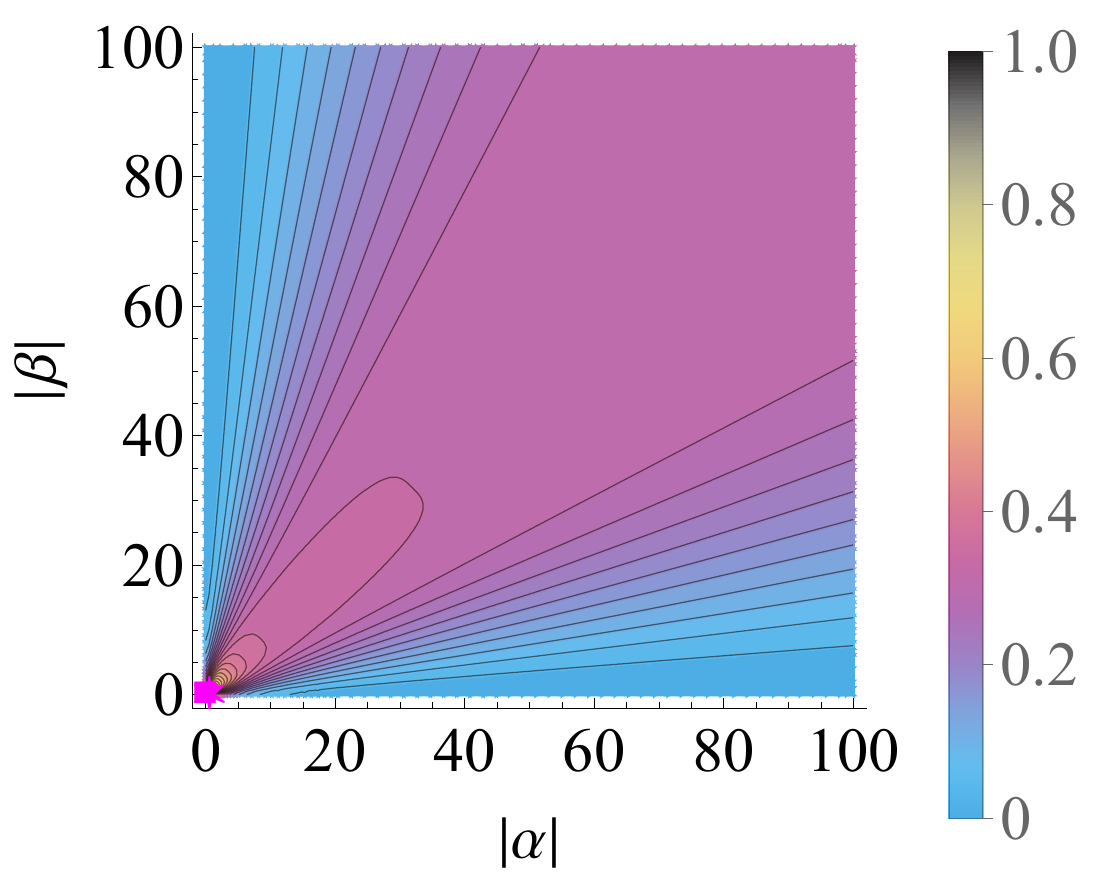}
 \caption{}
\label{purityTwo3CATd}
\end{subfigure}
\end{center}
\caption{Contour plots of (a) linear $\mathcal{L}_2^\infty$ and (b) von Neumann $\mathcal{S}_2^\infty$  entanglement entropies of the two-qutrit RDM $\rho_2(\3cat)$ of a $\rmu(3)$ Schr\"odinger cat \eqref{S3C} for $N\rightarrow\infty$, as a function 
of the phase-space coordinates $\alpha, \beta$ (they just depend on moduli). The meaning of the magenta curve is the same as in the Figure \ref{purityOne3CAT}. The asymptotic behaviour of (c) $\mathcal{L}_2^\infty$ and (d)  $\mathcal{S}_2^\infty$  for large values of $|\alpha|$ and $|\beta|$ displays isentropic curves $\theta=$constant, according to the expression of 
$\rho_2^\infty(\3cat)$ in Eq. \eqref{2RDM-alfabetainf}. }
\label{purityTwo3CAT}
\end{figure}

In Figure \ref{purityTwo3CATa}-\ref{purityTwo3CATb}, we 
represent contour plots of normalized linear  and von Neumann 
\be
\mathcal{L}_2^\infty=\frac{D^2}{D^2-1}(1-\tr((\rho_2^\infty)^2)),\quad \mathcal{S}_2^\infty=-\tr(\rho_{2}^\infty\log_{D^2} \rho_{2}^\infty),
\ee
pairwise entanglement entropies in the thermodynamic limit $N\rightarrow \infty$ for the two-qutrit RDM, $\rho_2^\infty(\3cat)$, of a $\rmu(3)$ Schr\"odinger cat \eqref{S3C}, as a function 
of the phase-space $\mathbb CP^{2}$ complex coordinates $(\alpha, \beta)$ [they just depend on the moduli]. As for the one-qutrit case, they attain their maximum 
value at the phase-space point $(\alpha,\beta)=(1,1)$ (``high coupling limit''); however, unlike the one-qutrit case, pairwise entanglement entropies do not attain the maximum value 
of 1 at this point, but $\mathcal{L}_2=5/6\simeq 0.833$ and $\mathcal{S}_2\simeq 0.623$ for large $N$. 
As already commented, variational (parity adapted spin coherent) approximations to the ground state of the  LMG 3-level atom model [discussed later in Section  \ref{LMGsec}] 
recover this maximum entanglement point  $(\alpha,\beta)=(1,1)$ at high interactions $\lambda\to\infty$, as can be seen in the already discussed stationary curve \eqref{critalphabeta}. 
In Figures \ref{purityTwo3CATc}  and \ref{purityTwo3CATd}, the asymptotic behavior for large $|\alpha|$ and $\beta|$ is shown, where 
contours of linear and von Neumann entropies coincide with the lines $\theta=$constant, according to the asymptotic behavior of $\rho_2^\infty(\3cat)$ in \eqref{2RDM-alfabetainf}.

We also plot in Figure \ref{InfDiag-2RDM} the information diagram for the  2-qutrit RDM of the $\3cat$ in the thermodynamic limit $N\rightarrow\infty$. It is clear that the $\Delta$ region is not completely filled; only the subregions $\Delta_2$ and $\Delta_3$ are partially filled, the reason being that  $\rho_2(\3cat)$ has rank $1$, $2$ or $4$. The  stationary curve $(\alpha_0(\lambda),\beta_0(\lambda))$ in Eq.   \eqref{critalphabeta} 
is also shown, with a behaviour similar to the case of the 1-qutrit RDM, with the difference that it ends near the maximally mixed RDM of rank 4, more precisely,  at the point $\left(\frac{1}{3},\frac{2}{9},\frac{2}{9},\frac{2}{9},0,0,0,0,0\right)$. It is important to notice that the stationary curve is most of the time at the inferior boundary of the set of 2-qutrit RDMs. This means that, from all 2-qutrit RDMs of the $\3cat$ with a given linear entropy, it has the minimum allowed value of von Neumann entropy. We conjecture that this is due to the variational character of
the ground state and the universality of the extremal states lying at the boundaries  of the region $\Delta$.

\section{Information diagrams and quantum phase transitions in Lipkin-Meshkov-Glick models of $3$-level identical atoms}
\label{LMGsec}

\noindent Now we apply the previous results to the study of QPTs of $D$-level Lipkin-Meshkov-Glick atom models. 
The standard case of  $D=2$ level atoms has already been studied in the literature (see e.g.  \cite{Calixto_2017}). We shall restrict ourselves to $D=3$ level atoms for practical calculations, 
although the procedure can be easily extended to general $D$. In particular, we propose the following LMG-type Hamiltonian 
\begin{equation}
H=\frac{\epsilon}{N}(S_{33}-S_{11})-\frac{\lambda}{N(N-1)}\sum_{i\not=j=1}^3 S_{ij}^2,\label{hamU3}
\end{equation}
written in terms of collective $\rmu(3)$-spin operators $S_{ij}$. Hamiltonians of this kind have already been proposed in the literature \cite{Kus,KusLipkin,Meredith,Casati,Saraceno} [see also 
\cite{nuestroPRE} for the role of mixed symmetry sectors in QPTs of multi-quDit LMG systems]. We place levels symmetrically about $i=2$, with  
intensive energy splitting per particle $\epsilon/N$. For simplicity, we consider equal interactions, with coupling constant $\lambda$, for atoms in different levels, 
and vanishing interactions for atoms in the same level (i.e., we discard interactions of the form $S_{ij}S_{ji}$). Therefore, $H$ is invariant under parity transformations $\Pi_j$ in \eqref{parityop}, 
since the interaction term scatters pairs of particles conserving the parity of the population $n_j$ in each level $j=1,\dots,D$. Energy levels have good parity, the ground state being an even state. 
We divide the two-body interaction in \eqref{hamU3} by the number of atom pairs $N(N-1)$ to make 
$H$ an intensive quantity, since we are interested in the thermodynamic limit $N\to\infty$. We shall see that parity symmetry is spontaneously broken in this limit. 

As already pointed long ago by Gilmore and coworkers \cite{GilmorePhysRevA.6.2211,RevModPhys.62.867}, coherent states constitute in general a powerful tool for rigorously studying the
ground state and critical properties of some physical systems in the thermodynamic limit. The energy surface associated to a  Hamiltonian density $H$  
is defined in general as the coherent state expectation value of the Hamiltonian density in the thermodynamic limit. In our case, the energy surface acquires the following form
\be
E_{(\alpha,\beta)}(\epsilon,\lambda)= \lim_{N\to\infty}\langle \zb|H|\zb\rangle= \epsilon\frac{  \beta  \bar{\beta }-1}{
\alpha  \bar{\alpha }+\beta  \bar{\beta }+1}
-\lambda\frac{ \alpha ^2 \left(\bar{\beta }^2+1\right)+\left(\beta ^2+1\right) \bar{\alpha }^2+\bar{\beta }^2+\beta ^2
}{\left(\alpha  \bar{\alpha }+\beta  \bar{\beta }+1\right)^2},\label{enersym}
\ee
where we have used 
the parametrization 
$\zb=(\alpha,\beta)$, as in eq. \eqref{S3C}, for $\rmu(3)$-spin coherent states $|\zb\ra$. Note that this energy surface is invariant under $\alpha\to-\alpha$ and  $\beta\to-\beta$, which is a consequence of 
the inherent parity symmetry of the  Hamiltonian \eqref{hamU3} and the transformation \eqref{parityCS} of $|\zb\ra$ under parity.

\begin{figure}[h]
\begin{center}
\begin{subfigure}[h]{0.48\textwidth}
 \centering
 \includegraphics[width=\textwidth]{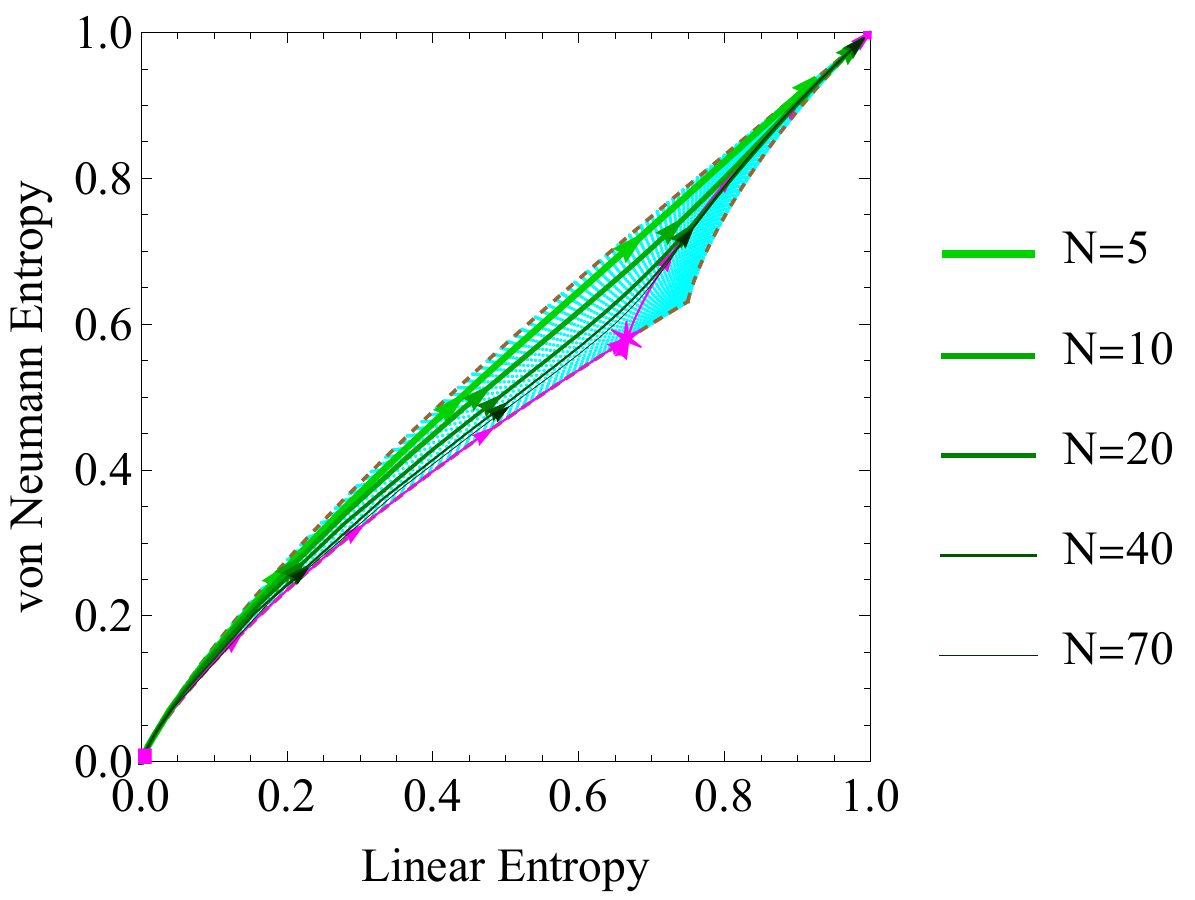}
 \caption{}
\label{InfDiag_NumCurvesa}
\end{subfigure}
\hspace{5mm}
\begin{subfigure}[h]{0.48\textwidth}
 \centering
 \includegraphics[width=\textwidth]{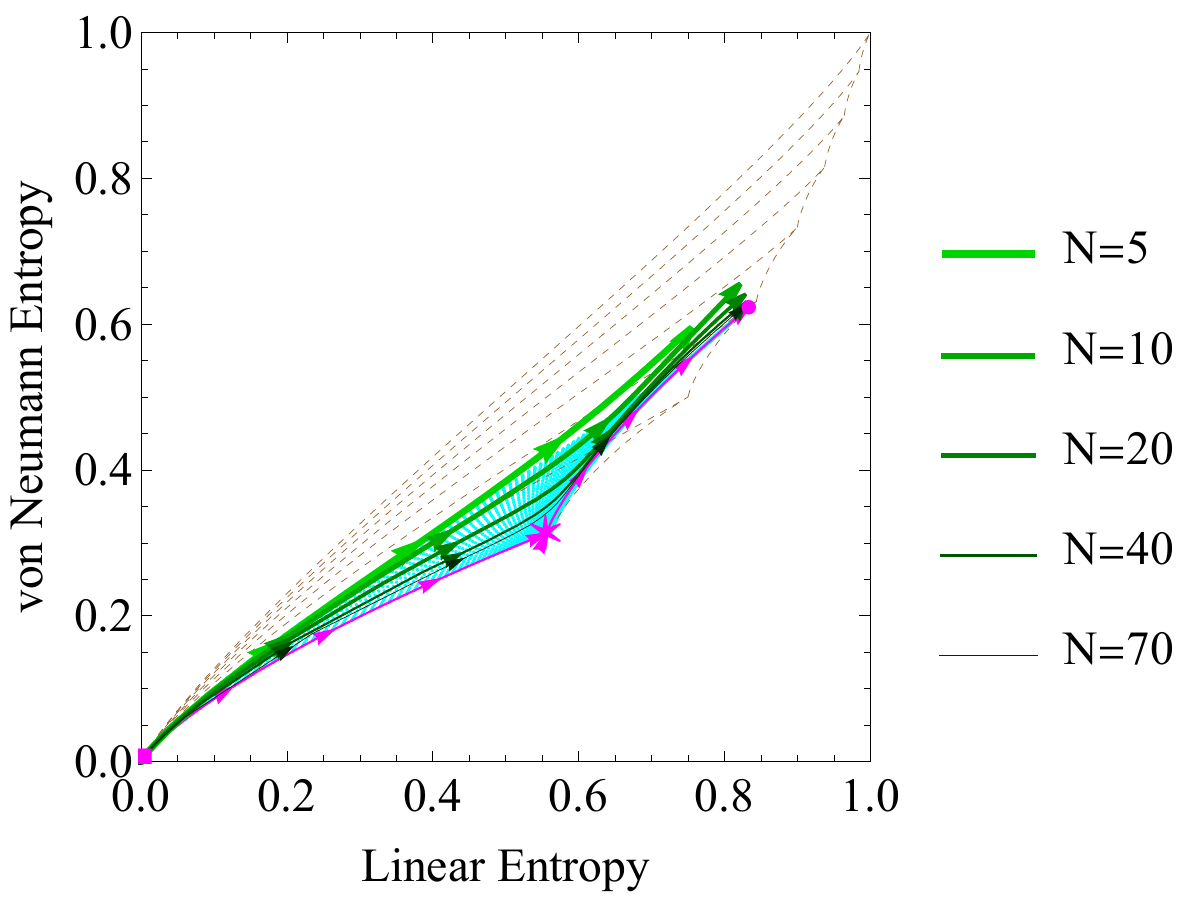}
 \caption{}
\label{InfDiag_NumCurvesb}
\end{subfigure}
\end{center}
\caption{Information diagram for the family of (a) 1-qutrit RDMs  and (b) 2-qutrit RDMs  for $\3cat$  in the limit $N\rightarrow\infty$, for all values of $|\alpha|$ and $|\beta|$, where the curves
of numerical RDMs, as a function of $\lambda$ for different values of $N$, has been added, as well as the analytical stationary curve for $N\rightarrow\infty$ (in magenta). Observe that, as $N$ grows, 
the numerical (green) curves approach the (magenta) analytical one.}
\label{InfDiag_NumCurves}
\end{figure}

The minimum energy 
\be E_0(\epsilon,\lambda)=\mathrm{min}_{\alpha,\beta\in \mathbb{C}}E_{(\alpha,\beta)}(\epsilon,\lambda)\label{minimieq}\ee
is attained at the stationary (real) phase-space values $\alpha_0^\pm=\pm \alpha_0$ and $\beta_0^\pm=\pm\beta_0$ with
\bea
\alpha_0(\epsilon,\lambda)&=&\left\{\begin{array}{lll}
 0, && 0\leq \lambda \leq \frac{\epsilon }{2}, \\
 \sqrt{\frac{2\lambda- \epsilon }{2 \lambda +\epsilon }}, && \frac{\epsilon }{2}\leq \lambda \leq \frac{3 \epsilon }{2}, \\
 \sqrt{\frac{2\lambda }{2 \lambda +3 \epsilon }}, && \lambda \geq \frac{3 \epsilon }{2},
\end{array}\right.\nonumber\\
\beta_0(\epsilon,\lambda)&=&\left\{\begin{array}{lll}
 0, & & 0\leq \lambda \leq  \frac{3 \epsilon }{2}, \\
 \sqrt{\frac{2 \lambda -3 \epsilon}{2 \lambda +3 \epsilon }}, & & \lambda \geq \frac{3 \epsilon }{2}. \end{array}\right. \label{critalphabeta}
\eea
In Figures  \ref{purityOne3CAT} and \ref{purityTwo3CAT}  we plotted (in magenta color) the stationary-point curve 
$(\alpha_0(\lambda),\beta_0(\lambda))$ on top of one- and two-qutrit entanglement entropies, noting that $(\alpha_0(\lambda),\beta_0(\lambda))\to (1,1)$ for  $\lambda\to\infty$ (high interactions). 
 Inserting \eqref{critalphabeta} into  \eqref{enersym} gives the ground state energy density at the thermodynamic limit 
\be
E_0(\epsilon,\lambda)=\left\{\begin{array}{lllr}
 -\epsilon,  && 0\leq \lambda \leq \frac{\epsilon }{2}, & \mathrm{(I)}\\
 -\frac{(2 \lambda +\epsilon )^2}{8 \lambda }, && \frac{\epsilon }{2}\leq \lambda \leq \frac{3 \epsilon }{2}, &  \mathrm{(II)} \\
  -\frac{4\lambda^2+3\epsilon ^2}{6 \lambda }, & &\lambda \geq \frac{3 \epsilon }{2}. &  \mathrm{(III)}\end{array}\right.\label{energysym}
\ee
Here we clearly distinguish three different phases: I, II and III, and two second-order QPTs at $\lambda^{(0)}_{\mathrm{I}\leftrightarrow\mathrm{II}}=\epsilon/2$ and 
$\lambda^{(0)}_{\mathrm{II}\leftrightarrow\mathrm{III}}=3\epsilon/2$, respectively, where $\frac{\partial^2E_0(\epsilon,\lambda)}{\partial\lambda^2}$ are discontinuous. In the stationary (magenta) curve $(\alpha_0(\lambda),\beta_0(\lambda))$ shown in Figures \ref{purityOne3CATa}, \ref{purityOne3CATb}, \ref{InfDiag-RDM}, \ref{purityTwo3CATa}, \ref{purityTwo3CATb}, and \ref{InfDiag_NumCurves},
the phase I corresponds to the origin $(\alpha_0,\beta_0)=(0,0)$ (square point), phase II corresponds to the horizontal part $\beta_0=0$ up to the star point, and phase III corresponds to $\beta_0\neq 0$. 

Note that the ground state is fourfold degenerated in the thermodynamic limit since the four $\rmu(3)$-spin coherent states $|\zb_0^{\pm\pm}\ra=|\pm\alpha_0,\pm\beta_0\ra$ have the same energy density $E_0$. These four coherent states
are related by parity transformations  and, therefore, parity symmetry is spontaneously broken in the thermodynamic limit. In order to have good variational states for finite $N$,  
to compare with numerical calculations, we have two possibilities: 1) either we use the $\3cat$ \eqref{S3C} as an ansatz for the ground state, minimizing $\langle\3cat|H|\3cat\rangle$, or 2) we restore 
the parity symmetry of the coherent state $|\alpha_0,\beta_0\ra$ for finite $N$ by projecting on the even parity sector. Although the first possibility offers a more accurate variational approximation to the ground state, 
it entails a more tedious numerical minimization than the one already obtained in \eqref{minimieq} for $N\to\infty$. Therefore, we shall use the second  possibility which, despite being less accurate, it is 
straightforward and good enough for our purposes. That is, we shall use the $\3cat$  \eqref{S3C}, evaluated at $\alpha=\alpha_0$ and $\beta=\beta_0$, as a variational approximation 
$|\3cat_0\ra$ to the numerical (exact) ground state $|\psi_0\ra$ for finite $N$.

Let us apply the tools developed in previous sections to this model and draw the main conclusions. Firstly,  in Figure \ref{InfDiag_NumCurvesa} and \ref{InfDiag_NumCurvesb},  we have added to the information diagrams for 1 and 2 qutrits RDMs already shown in Figure \ref{InfDiag-RDM}, the  curves (as a function of $\lambda$) of the numerically computed ground states of the 3-level LMG model for different values on $N$ (in green colors), together with the already shown  analytical variational curve (in magenta) $(\alpha_0(\lambda),\beta_0(\lambda))$ for $N\rightarrow\infty$. We can conclude  that they do not lie in the inferior part of the region $\Delta$, as the variational one, but  as $N$ grows the numerical curves approach the analytical one.

\begin{figure}[h]
\begin{center}
 \includegraphics[width=12cm]{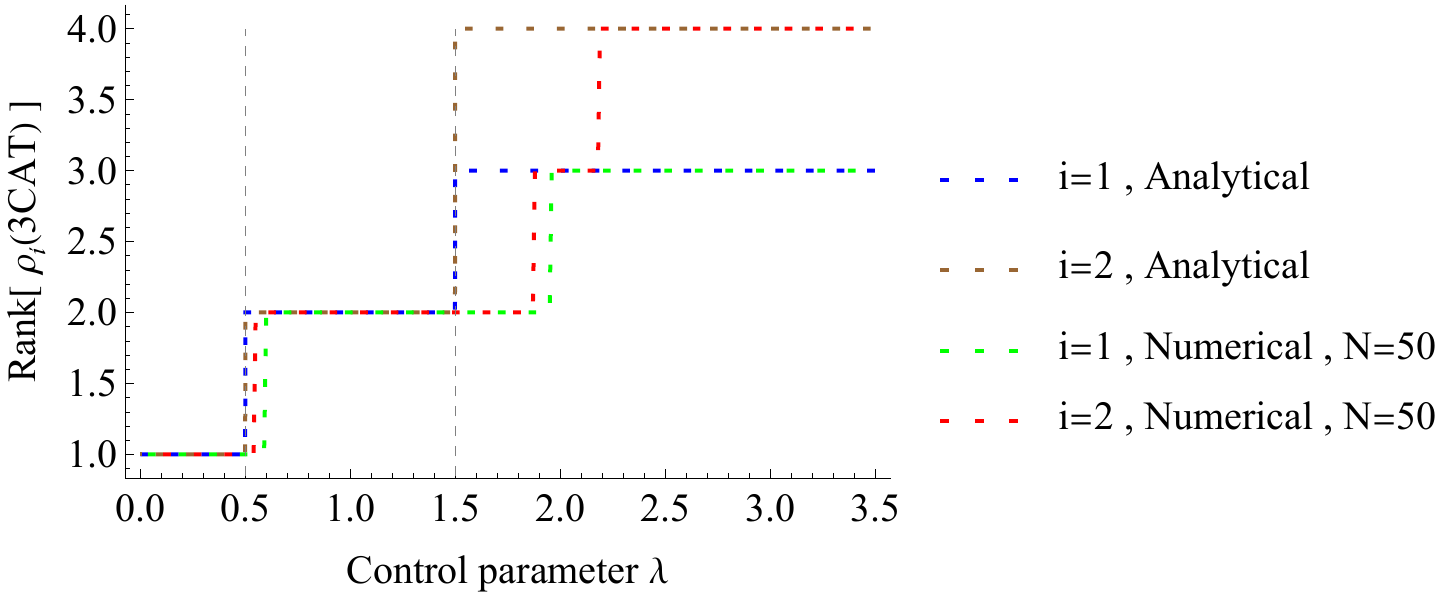}
 \caption{Plot of rank of 1-quDit and 2-quDit RDMs along the stationary curve both for analytical (variational, $N\rightarrow\infty$) and numerical ($N=50$) solution of the 3-level LMG model as a function of $\lambda$ (in $\epsilon=1$ units).}
\label{RankNumericalPlusAnalytical}
\end{center}
\end{figure}

Secondly, suggested by the results about the rank of 1 and 2 quDits RDMs of Section \ref{Entanglement}, we plot in Figure \ref{RankNumericalPlusAnalytical} the rank of the RDMs as a function of $\lambda$ for both variational ($N\rightarrow\infty$) and numerical ($N=50$) solutions for the ground state of Hamiltonian \eqref{hamU3}. The QPT critical points $\lambda^{(0)}_{\mathrm{I}\leftrightarrow\mathrm{II}}=\epsilon/2$ and $\lambda^{(0)}_{\mathrm{II}\leftrightarrow\mathrm{III}}=3\epsilon/2$, are clearly marked in the case of the variational curve, with a jump from rank 1 to rank 2 at $\lambda=\epsilon/2$ and another jump from rank 2 to rank 4 (3 in the case of 1 qutrit RDMs) at $\lambda=3\epsilon/2$. In the case of the numerical curve, where a small threshold has been applied to the eigenvalues to suppress spurious oscillations, the first jump continues to be at $\lambda\simeq 0.5$, whereas the second jump takes place at slightly larges values of $\lambda=1.5$ (in $\epsilon=1$ units). This behaviour is the same as with other precursors of QPTs, like susceptibility of fidelity in the 3-level LMG model \cite{nuestroPRE}.

From this, it is clear that the rank of the RDMs is a good precursor of a QPT, with the advantage of being a discrete parameter.

\section{Conclusions}
\label{conclusions}

In this paper we have used an information-theoretic tool like the information diagrams to extract qualitative information about the quDit entanglement (and rank) of parity adapted $\rmu(D)$-spin coherent states ($\dcat$s) using one- and two-quDit reduced density matrices, and we have applied it to the study of atom entanglement in the  ground state (both variational, in the $N\rightarrow\infty$, and numerical, with finite $N$)  of the 3-level atom LMG model.

We have shown how the allowed region $\Delta$ of information diagrams is completely filled in the case of one-qutrit RDMs, while only the lower part of it is partially filled in the case of two-qutrits RDMs. This indicates that the maximum pairwise (2-qutrit) entanglement attained in a $\3cat$ state is smaller that the maximum one corresponding to a maximally mixed RDM or order $3^2$. We have already seen that this maximally  entangled $\3cat$ is attained for the values $(\alpha,\beta)=(1,1)$ (or $\alpha=1$ for $D=2$), and these are precisely the values obtained for the variational analytical approximation to the ground state of a 3-level LMG model in the high coupling regime.

In addition, we have shown that the variational  curve $(\alpha_0(\lambda),\beta_0(\lambda))$ practically all the time lies in the inferior part of the information diagram subregion filled by all $\3cat$ states. We conjecture that this is due to the variational character of these states (minimum of the energy surface \eqref{enersym}) and the universality character of the extremal states lying at the boundary of the region $\Delta$.

Information diagrams also provide qualitative information about the rank of the RDMs. This has motivated us to study with detail their rank for different values of the parameters $\alpha$ and $\beta$ of $\3cat$ states (see Section \ref{Entanglement}),
indicating that the one- and two-quDit RDMs have in general lower ranks than the maximal rank allowed by the corresponding dimension. Focusing on the 
variational analytic curve $(\alpha_0(\lambda),\beta_0(\lambda))$, and in the numerical solution for the ground state for finite $N$,   Figure \ref{RankNumericalPlusAnalytical}  shows that the rank of one- and two-qutrit RDMs has jumps precisely at the points where QPTs occurs (or near these values in the numerical finite $N$ case). Therefore the rank can be used as a discrete precursor of a QPT in the LMG model, but this conclusion  can be probably extended to other critical models.

All these results motivate us to further study the application of information diagrams and rank of RDMs to other  parity adapted $\rmu(D)$-spin coherent states,  but with different parity character. Here we have restricted ourselves to the even case, but remember that there are $2^{D-1}$ different parity adapted $\rmu(D)$-spin coherent states, the even one just being a particular case. For example, odd parity cat states (for $D=2$) are known to be well suited as variational states to approximate excited states in, for example, the Dicke model of superradiance \cite{PRA2011-Octavio-Superradiance}.

Since the rank of a RDM is equal to the Schmidt number, by the Schmidt decomposition theorem (see, for instance 
\cite{PRL2013-Huber-SchmidtNumber}), it would be interesting to study with detail the Schmidt decomposition of parity adapted $\rmu(D)$-spin coherent states (not only of the even one, but for all $2^{D-1}$ parity invariant states) when we extract 1, 2, or in general $M$ quDits, and find the basis realizing the Schmidt decomposition in the larger factor.

\section*{Acknowledgments}
 We thank the support of the Spanish MICINN  through the project PGC2018-097831-B-I00 and  Junta de Andaluc\'\i a through the projects SOMM17/6105/UGR, UHU-1262561 and FQM-381. 
 AM thanks the Spanish MIU for the FPU19/06376 predoctoral fellowship.

\bibliography{/home/guerrero/MEGA/Genfimat/Bibliografia/bibliografia.bib}

\end{document}